\documentclass[manuscript, screen]{acmart}
\AtBeginDocument{%
  \providecommand\BibTeX{{%
    \normalfont B\kern-0.5em{\scshape i\kern-0.25em b}\kern-0.8em\TeX}}}

\setcopyright{none}
\renewcommand\footnotetextcopyrightpermission[1]{} 

\usepackage{threeparttable}
\usepackage{subfigure}
\usepackage{multirow}
\usepackage{amsmath}
\usepackage{commath}
\usepackage{multirow}
\usepackage{booktabs}
\usepackage{enumitem}
\usepackage{lscape}
\usepackage{amsfonts}
\usepackage{siunitx}
\usepackage{makecell}  
\usepackage{geometry}  
\usepackage{array}     
\usepackage{tabularx}  
\usepackage{ragged2e}
\usepackage{color}
\newif\ifMARKED
\MARKEDtrue  

\ifMARKED
    
\else
    
\fi

\ifMARKED

\else
    
\fi

\begin{document}

\pagestyle{plain} 
\title{PocketPPD: Screening for Postpartum Depression Risk Using Passive Smartphone Sensing}



\author{Jia Tang}
\affiliation{%
 \institution{The University of Tokyo}
 \city{Tokyo}
 \country{Japan}
}
\email{jiatang@mcl.iis.u-tokyo.ac.jp}

\author{Helinyi Peng}
\affiliation{%
 \institution{The University of Tokyo}
 \city{Tokyo}
 \country{Japan}
}
\email{penghly@outlook.com}

\author{Akihito Taya}
\affiliation{%
 \institution{The University of Tokyo}
 \city{Tokyo}
 \country{Japan}
}
\email{taya-a@iis.u-tokyo.ac.jp}

\author{Kaoru Sezaki}
\affiliation{%
 \institution{The University of Tokyo}
 \city{Tokyo}
 \country{Japan}
}
\email{sezaki@iis.u-tokyo.ac.jp}

\author{Daisuke Nishi}
\affiliation{%
 \institution{The University of Tokyo}
 \city{Tokyo}
 \country{Japan}
}
\email{d-nishi@m.u-tokyo.ac.jp}

\author{Anind K Dey}
\affiliation{%
 \institution{University of Washington}
 \city{Seattle}
 \country{United States}
}
\email{anind@uw.edu}

\author{Yuuki Nishiyama}
\affiliation{%
 \institution{The University of Tokyo}
 \city{Tokyo}
 \country{Japan}
}
\email{nishiyama@csis.u-tokyo.ac.jp}

\begin{abstract}

Postpartum depression (PPD) is a serious perinatal mental health condition affecting approximately 20\% of new mothers worldwide.
Common screening approaches for PPD, such as self-report questionnaires and active digital logs, rely heavily on user input and thus impose a substantial burden on participants, limiting their feasibility for long-term use.
Recent passive mobile sensing (PMS) approaches have enabled low-burden detection of depressive symptoms using machine learning methods with multi-modal sensor data from off-the-shelf mobile devices including smartphones.
However, the postpartum period entails distinct behavioral patterns, raising uncertainty about whether sensing-based indicators for general depression and mental disorders generalize to PPD.
To address this gap, we propose PocketPPD, a PMS-based PPD screening method that detects PPD risk using maternal contextual features, such as disruptions in behavioral rhythms and shifts in stability, collected through a smartphone.
In our exploratory four-week feasibility study with 61 postpartum women, the PMS-only model achieved an AUC of 0.75, while the best-performing model, integrating PMS-oriented data and self-report features, achieved an AUC of 0.83. 
Moreover, we find that morning and late-night routine volatility ranks among the top digital biomarkers, dynamically moderated by maternal contexts such as infant developmental stage and employment status.
This work provides empirical evidence for low-burden PPD risk screening and our findings lay the groundwork for continuous perinatal mental health monitoring.

\end{abstract}




\maketitle

\section{Introduction}
Postpartum depression (PPD) is one of the most common and debilitating perinatal mental health conditions, with a global pooled prevalence estimated at approximately 20\% across 46 countries~\cite{FishWilliamson2023}. In Japan, approximately 15\% of women are affected by PPD within the first month after childbirth~\cite{Tokumitsu2020,Takeda2020,Suzuki2024}. 
The impact of PPD extends beyond the mother, leading to adverse financial consequences~\cite{Luca2020} and significantly increasing the risk of emotional disorders in children and depressive symptoms in partners~\cite{Kavanaugh2006, Duan2020}. Despite its severe consequences, PPD remains substantially underdiagnosed and undertreated, with reported treatment rates as low as 15\%~\cite{Hutchens2020}. This clinical gap is largely attributable to the structural limitations of current screening protocols. 
The Edinburgh Postnatal Depression Scale (EPDS)~\cite{epds} is the most widely used screening tool for PPD worldwide. However, it relies on self-reports and is typically administered at discrete time points (e.g., 1 month postpartum), which fails to capture the dynamic symptom fluctuations during the perinatal period. While traditional gold standards and active digital sensing (e.g., daily mood logs and Ecological Momentary Assessments: EMA) can achieve strong screening performance when combined with baseline clinical scales~\cite{Hahn2021, Allen2024}, they demand continuous user participation. For new mothers experiencing significant fatigue and time scarcity, this imposes a considerable burden, frequently leading to data missingness and failing to capture real-time symptom fluctuations unobtrusively. There is therefore an urgent clinical need for a low-burden, continuous passive screening approach.

\emph{Passive mobile sensing} (PMS) is a method that leverages continuous, unobtrusive signals from ubiquitous mobile devices, including smartphones and wearables such as smartwatches, to infer behavioral and physiological states, thereby alleviating the burden and allowing for continuous mental health monitoring. PMS has been widely applied to screen depression in the general population~\cite{Zou2023,Wang2018,29}, and a smaller body of work has begun to extend the paradigm to PPD using either wearables~\cite{Hurwitz2024, Slyepchenko2022} or smartphones~\cite{Fransson2022}. However, neither line of work transfers directly---in methodology or in empirical findings---to the setting we target, namely fine-grained, smartphone-only PPD screening. The mismatch arises along three dimensions.

First, methodologies and digital biomarkers established in general-population PMS studies do not generalize to postpartum mothers, who constitute a distinct clinical population shaped by behavioral patterns and contextual factors (e.g., infant developmental stages, childcare demands, and family support structures) that differ substantially from those of the general population. As a result, biomarkers that interpret reduced activity volume or a smaller social space as lethargy in the general population can be misleading in the postpartum period, where such patterns may instead reflect caregiving demands and the broader maternal context.

Second, prior PMS-based PPD screening methods that achieve the strongest performance rely on wearable devices (e.g., Fitbit) to capture precise physiological signals~\cite{Hurwitz2024, Slyepchenko2022}, introducing hardware costs and usability barriers that limit population coverage. For instance, while smartwatch ownership in Japan stands at approximately 16.2\%~\cite{docomo2025}, smartphone penetration in the country exceeds 98\%, highlighting the coverage gap between wearable- and smartphone-based approaches.

Third, existing smartphone-only PPD studies have largely been confined to coarse temporal aggregation and a narrow GPS-derived feature set---such as the monthly median distance from home---and have reported weak or inconsistent associations with depressive symptoms~\cite{Fransson2022}, leaving the feasibility of fine-grained smartphone-only PPD screening largely unaddressed.

\begin{figure}[tb]
\centering
\includegraphics[width=\linewidth]{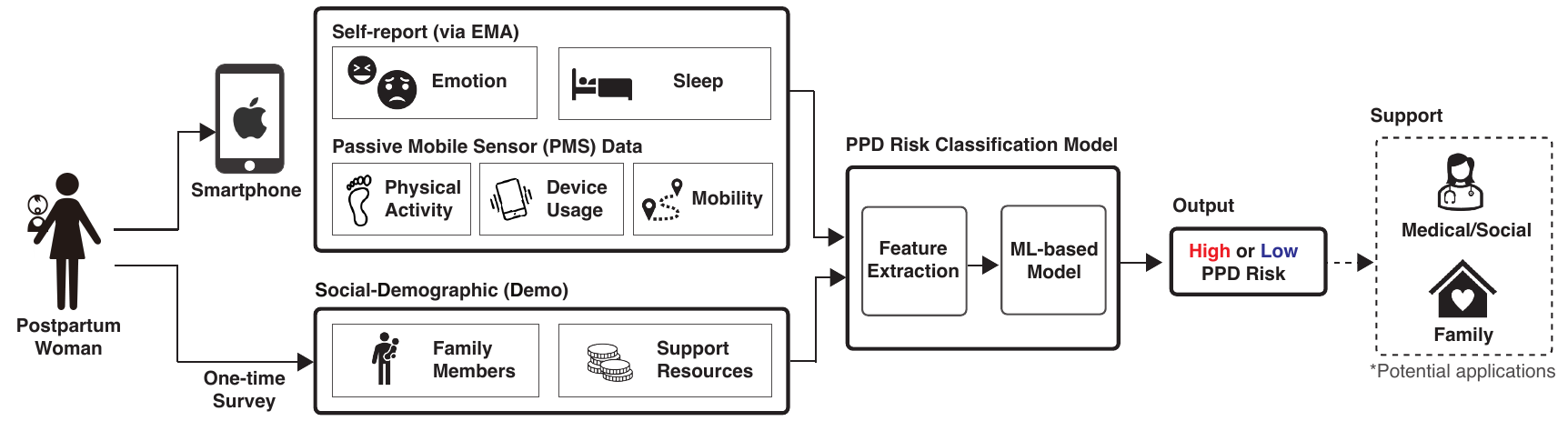}
\caption{Overview of the proposed smartphone-based PPD risk screening method and its data processing pipeline. A postpartum participant provides daily activity data via a smartphone, including (i) active self-reports (by EMA), (ii) passively collected hardware and software sensor data (e.g., physical activity, device usage, and mobility), and (iii) Social-Demographic information. These inputs are transformed through feature extraction and fed into ML-based models to screen binary PPD risk.}
\label{fig:overview}
\end{figure}

To address these challenges, we developed PocketPPD, a smartphone-only PMS-based PPD screening method that detects EPDS-defined binary PPD risk as shown in Figure~\ref{fig:overview}. 
The method is built on two key design principles.
First, we extracted a feature set for a PPD risk detection algorithm from smartphone-based daily behavior data, grounded in the premise that the most informative digital biomarkers for PPD lie in disruptions and shifts in behavioral rhythms tied to the maternal context. Specifically, we engineer features at fine temporal granularity and link them to maternal contextual factors, then apply interpretable machine learning under a cross-validation protocol; post-hoc Linear Mixed Models (LMMs) further quantify how behavioral rhythms interact with contextual moderators such as infant developmental stage and employment status.
Second, the method is designed around a \textit{passive-primary, active-auxiliary} principle: users contribute continuous smartphone PMS streams as the primary signal, supplemented by only sparse, lightweight self-reports---a one-time baseline survey for socioeconomic status (SES) and a brief morning EMA---so as to minimize compliance demands.


To evaluate this method, we conducted a four-week study with 61 postpartum mothers in Japan to screen for EPDS-defined binary PPD risk. 
Our experimental results and source-contribution ablation confirm that passive mobile sensing (PMS) carries the central screening signal on its own, while lightweight active streams contribute a meaningful incremental gain. Moreover, we find that morning and late-night routine volatility offers strong predictive power and is dynamically moderated by maternal contexts, such as infant developmental stage and employment status. Finally, we demonstrate our method's robustness across different EPDS thresholds and discuss how these digital biomarkers diverge from general-depression phenotypes.

This work provides empirical evidence for smartphone only continuous PPD risk screening and our findings lay the groundwork for continuous perinatal mental health monitoring. Our main contributions are as follows:
\begin{itemize}
    \item \textbf{A wearable-free smartphone-only PPD risk screening method.}
    We develop and feasibility-test PocketPPD, a smartphone-only PMS-based method that detects EPDS-defined PPD risk without relying on wearable devices. Fine-grained temporal feature engineering combined with interpretable machine learning achieves an AUC of 0.83 in a four-week study with 61 postpartum mothers in Japan.

    \item \textbf{Empirical evidence for passive-primary, active-auxiliary design.}
    We validate a design strategy in which smartphone PMS streams serve as the primary signal, supplemented by lightweight self-reports to minimize user burden. A source-contribution ablation confirms that PMS alone achieves AUC 0.75, while sparse EMA and Social-Demographic inputs provide a meaningful incremental gain.

    \item \textbf{PPD-specific digital biomarkers tied to maternal context.}
    We identify morning and late-night routine volatility as strong predictors of PPD risk, moderated by infant developmental stage and employment status. These markers diverge from general-depression phenotypes, reflecting the unique demands of the postpartum period.
\end{itemize}

\section{Related Work}
We review prior work along three lines: conventional PPD screening, behavioral risk markers from depression and PPD studies, and machine-learning approaches for perinatal depression detection. The main distinction is that general depression sensing has explored rich passive signals and temporal modeling, whereas PPD research has more often relied on active self-report, wearables, or coarse smartphone aggregates.

\begin{table*}[tb]
\centering
\caption{Comparison of Related Work on Perinatal Depression Predictions}
\label{tab:ppd_research}
\scriptsize
\setlength{\tabcolsep}{3pt}
\renewcommand{\arraystretch}{1.45}
\newcolumntype{Y}{>{\raggedright\arraybackslash}X}

\begin{tabularx}{\textwidth}{
@{} >
{\raggedright\arraybackslash}p{3.25cm} >{\raggedright\arraybackslash}p{1.65cm} >{\raggedright\arraybackslash}p{1.9cm} c > 
{\raggedright\arraybackslash}p{2.05cm} >
{\raggedright\arraybackslash}p{1.85cm} >
{\hsize=1.0\hsize}Y 
@{}
}
\toprule
\textbf{Theme \& Gap} &
\textbf{Study} &
\textbf{Data Modality \& Collection} &
\textbf{Term} &
\textbf{Extracted Features} &
\textbf{Modeling Strategy} &
\textbf{Target \& Performance} \\
\midrule

\raggedright
{\bf Active Digital Phenotyping}
\newline
$\bullet$ Requires continuous active participation\newline
$\bullet$ High recruitment and longitudinal cost
& Hahn et al. \cite{Hahn2021}
& Web Platform \newline \textit{Self-report}
& Postpartum
& $\bullet$ EPDS, mood/stress scores \& change rates
& \textbf{Global Model} \newline $\bullet$ LR
& \textbf{Depression Risk} \newline Bal.\ Acc: 93\% \newline Specificity: 99\% \\
\addlinespace

& Zhong et al. \cite{Zhong2022}
& Smartphone \newline \textit{Self-report}
& Pregnancy
& $\bullet$ Previous EPDS, WHO-5$^\dagger$, anxiety history, stress
& \textbf{Global Model} \newline $\bullet$ LR \newline $\bullet$ Decision-level fusion
& \textbf{Depression Risk} \newline AUC: 0.82 \newline BAC: 0.75 \\
\addlinespace

& Krishnamurti et al. \cite{Krishnamurti2023}
& Smartphone \newline \textit{Self-report}
& Pregnancy
& $\bullet$ Linguistic$^\ddagger$, daily mood
& \textbf{Global Model} \newline $\bullet$ LASSO Regression
& \textbf{Depression Risk} \newline AUC: 0.84 \\
\addlinespace

& Allen et al. \cite{Allen2024}
& Smartphone \newline \textit{Self-report}
& Pregnancy
& $\bullet$ Daily mood, acute symptoms, text sentiment
& \textbf{Global Model} \newline $\bullet$ Stepwise LASSO
& \textbf{Depression Risk} \newline AUC: 0.82--0.83 \\

\midrule

\raggedright
{\bf Wearable-based Physiological \& Rhythm Sensing}
\newline
$\bullet$ Hardware and wearability burden
& Slyepchenko et al. \cite{Slyepchenko2022}
& Wearable Device \newline (Actigraph) \newline \textit{PMS}
& Postpartum
& $\bullet$ Nighttime activity, circadian quotient, sleep-related features
& \textbf{Global Model} \newline $\bullet$ GEE $^\ast$
& \textbf{Depressive Symp.} \newline Circadian rhythms were stronger predictors than sleep \\
\addlinespace

& Hurwitz et al. \cite{Hurwitz2024}
& Wearable Device \newline (Fitbit) \newline \textit{PMS}
& Postpartum
& $\bullet$ Calories BMR, average HR, steps, worn-to-sleep frequency
& \textbf{Personalized} \newline $\bullet$ Random Forest 
& \textbf{PPD Period Ident.} \newline mAUC: 0.85 \newline F1: 0.81 \\

\midrule

{\bf Coarse-grained Smartphone Passive Sensing}
\newline
$\bullet$ Coarse temporal resolution
\newline
$\bullet$ Single GPS-derived feature
& Fransson et al. \cite{Fransson2022}
& Smartphone \newline \textit{PMS}
& Pregnancy
& $\bullet$ Continuous GPS tracking: Median distance from home
& \textbf{Global Model} \newline $\bullet$ LR \& Linear Regression $^\ast$
& \textbf{Depressive Symp.} \newline Mobility not significant \\

\midrule

{\bf PocketPPD: Passive-primary, Active-auxiliary Smartphone Sensing}
\newline
$\bullet$ Passive sensing as primary signal
& Ours
& Smartphone \newline \textit{PMS (+ Self-report)}
& Postpartum
& $\bullet$ PMS: physical activity, mobility, phone usage \newline $\bullet$ EMA: emotion, sleep; Demo: social-demographics
& \textbf{Global Model} \newline $\bullet$ XGBoost
& \textbf{PPD Risk} 
\newline
[\textit{PMS-only Model}]: 
\newline AUC: 0.754, Sens.: 0.787 \newline
[\textit{All Sources Model}]:
\newline AUC: 0.832, Sens.: 0.821 
\\

\bottomrule
\end{tabularx}
\vspace{1ex}

\raggedright
\textit{Note:} PMS, LR, and Demo denote Passive Mobile Sensing, Logistic Regression, and Socio-demographics, respectively. 
$^\ast$Did not use a machine learning model; GEE denotes Generalized Estimating Equations.
$^\dagger$WHO-5 refers to the World Health Organization Five Well-Being Index.
$^\ddagger$Linguistic features include word embeddings and related text-derived markers.
\end{table*}

\subsection{Postpartum Depression Screening}

PPD is typically assessed through structured clinical interviews and screening instruments such as the Edinburgh Postnatal Depression Scale (EPDS)~\cite{epds}, with follow-up assessment using instruments such as the Hamilton Depression Rating Scale~\cite{Hamilton1960}. These protocols are clinically established, but they are difficult to use for continuous monitoring. In-clinic assessments are costly and time-consuming, and many studies therefore rely on cross-sectional designs or measurements repeated at intervals as sparse as monthly visits~\cite{Hahn2021}.
Moreover, screening at a single postpartum time point, or only in later windows such as 8--32 weeks, can miss the dynamic symptom trajectories that unfold after childbirth~\cite{Tortajada2009, JimenezSerrano2015}. These limitations motivate low-burden sensing approaches that can complement, rather than replace, clinical screening.

\subsection{Behavioral Risk Patterns}
Digital phenotyping studies of general depression consistently link depressive symptoms with changes in mobility, activity rhythms, and social communication. A shrinking life space---operationalized through lower location entropy, increased homestay, and fewer visited places---is a frequently reported mobility marker~\cite{Zhang2022, Sun2023}; decreased physical activity and disrupted rhythms have also been repeatedly reported~\cite{Wang2018, mehrotra2018Using, wang2022FirstGen}. Communication patterns provide another signal: depression severity has been associated with shorter calls, fewer sent messages, and lower diversity of communication-app use~\cite{Liu2023, tlachac2024Symptom}.

These markers are useful starting points for PPD, but they cannot be transferred naively. Postpartum routines are shaped by childcare demands, infant developmental stage, household support, and employment status; consequently, low mobility or irregular nighttime activity may reflect caregiving context rather than depressive withdrawal. PPD-specific studies reinforce this contextual dependence. Family or partner support is a salient risk factor, whereas static demographic variables alone provide limited predictive power. In perinatal populations more broadly, text and social-media studies further suggest that linguistic markers of support and self-focus, such as first-person plural or singular pronouns, can relate to later depressive symptoms~\cite{Krishnamurti2023}.
Wearable studies have highlighted circadian and physiological rhythms, including circadian quotient, heart rate, and light exposure, as informative PPD indicators~\cite{Slyepchenko2022, Hurwitz2024}.
Objective sleep metrics also contribute: shorter total sleep time and greater sleep fragmentation during the perinatal period have been associated with elevated depressive symptoms~\cite{Pitsillos2022,Micheletti2020}.
In contrast, smartphone-only PPD studies using coarse GPS summaries, such as monthly median distance from home, have found weak or inconsistent associations~\cite{Fransson2022}. Taken together, prior work suggests that PPD sensing should emphasize behavioral rhythm, stability, and context, rather than only absolute activity volume.

\subsection{Machine Learning-based Depression Detection}
Machine-learning studies of general depression have established a broad technical repertoire for passive sensing. Large cohorts and longer observation windows~\cite{Bai2021, Lee2023a, Zhang2022, Nepal2024} support multimodal fusion of smartphone logs and wearable data~\cite{Chikersal2021, Lee2023a, Xu2021, lu2018Joint}, temporal slicing~\cite{Chikersal2021}, circadian feature extraction~\cite{Lee2023a}, and sequential models such as RNNs or GRU-Ds~\cite{Zou2023, salekin2018Weakly}. These studies demonstrate the feasibility of sensing-based depression detection, but their scale, sensing burden, and general-population assumptions do not directly match the constraints of postpartum cohorts.

Perinatal and postpartum depression screening remains more constrained. Many high-performing systems use active reports as the primary signal, including repeated clinical scales, mood and stress ratings, daily check-ins, or free-text journals processed with NLP~\cite{Hahn2021, Zhong2022, Krishnamurti2023, Allen2024}. These approaches can achieve strong AUC or balanced accuracy, but they require sustained participation over long periods. Passive approaches reduce active burden, and the strong PPD results have often depended on wearable-derived physiological signals~\cite{Hurwitz2024}, for instance, reached an mAUC of 0.85 with personalized random-forest models on Fitbit-derived heart rate and activity data. Slyepchenko et al.~\cite{Slyepchenko2022} similarly found that actigraphy-derived circadian rhythm metrics outperformed sleep duration measures in associating with depressive symptoms.
Smartphone-only passive sensing is more scalable, yet existing work has largely used coarse temporal aggregation or a narrow GPS-derived feature set~\cite{Fransson2022}. Table~\ref{tab:ppd_research} summarizes these branches and their trade-offs.

Overall, prior work has not yet combined three properties that are important for scalable PPD screening: reliance on ubiquitous smartphones rather than dedicated wearables, fine-grained temporal resolution, and features designed around postpartum behavioral rhythms and context. PocketPPD addresses this gap on all three fronts. It extracts temporally sliced smartphone biomarkers focused on behavioral rhythm and stability, interprets them through maternal context, and pairs these dense passive streams with sparse, lightweight self-reports.

\section{Motivation and Approach}

No prior approach simultaneously satisfies the three properties required for continuous, scalable PPD screening:
(i)~a passive-primary design that minimizes active compliance burden while relying on ubiquitous smartphones, (ii)~fine-grained feature design grounded in postpartum behavioral rhythms, and (iii)~contextual grounding in the postpartum period rather than reliance on general-depression phenotypes. Building on these gaps, we formulate three research questions and describe the corresponding methodological approach for each.

\textbf{From Compliance Trade-off to Passive-Primary Design.}
Active-primary designs achieve strong screening performance but accumulate weeks of self-report burden, while fully passive designs forgo contextual signals~\cite{Hahn2021, Allen2024}. 
PocketPPD navigates this trade-off through a \textit{passive-primary, active-auxiliary} configuration: dense PMS streams as the primary signal, supplemented by a one-time Social-Demographic survey and a brief morning EMA. We quantify each source's contribution through an ablation that enumerates all combinations of PMS, EMA, and Social-Demographic inputs.

\begin{itemize}
\item \textbf{RQ1}: How well can PPD risk be screened from passive smartphone data alone, and what incremental gain do supplementary self-report sources provide?
\end{itemize}

\textbf{From Absolute Volume to Behavioral Rhythm and Volatility.}
Prior smartphone-only PMS studies relied on coarse absolute-volume metrics (e.g., monthly GPS distance from home) and reported weak or inconsistent associations~\cite{Fransson2022}. These failures suggest that the relevant signal lies not in \textit{how much} a mother moves, but in \textit{when} and \textit{how regularly} she does so. To operationalize this shift from volume to rhythm, we engineer features at fine temporal granularity (four 6-hour windows $\times$ workday/off-day context), extract circadian rhythm metrics (Interdaily Stability, Intradaily Variability, M10/L5 phase), and apply interpretable machine learning combined with SHAP-based feature importance and LMM-based statistical validation to surface robust, clinically interpretable markers.

\begin{itemize}
\item \textbf{RQ2}: What behavioral rhythm and volatility markers derived from fine-grained smartphone data are associated with PPD risk, and how do maternal demographic factors modulate these associations?
\end{itemize}

\textbf{From Population Phenotypes to Maternal Context.}

General-depression markers such as reduced mobility and social disengagement~\cite{Chikersal2021, Xu2021, Zhang2022} may not transfer to postpartum mothers, where the same patterns can instead reflect caregiving demands, infant developmental stage, or household support.
We complement LMM main-effect analysis with moderation tests on maternal contextual factors (infant age, employment status, family support) and interpret the identified digital biomarkers against the general-depression literature.
\begin{itemize}
\item \textbf{RQ3}: How do the identified PPD-specific digital biomarkers diverge from general-depression phenotypes, and to what extent are they moderated by maternal contextual factors such as infant developmental stage and employment status?
\end{itemize}

\section{Study}
\label{sec:dataset}
We conducted a four-week observational study with 61 postpartum mothers in Japan to evaluate the proposed PPD screening method. The study combines continuous smartphone PMS, daily morning EMA, and weekly EPDS labels through an AWARE-based collection platform, yielding 227 valid weekly observations across three independent rounds.

\subsection{Data Collection Platform}
We developed a data-collection platform based on the AWARE Framework \cite{15aware1} to passively capture PMS data from personal smartphones. The application also facilitated the collection of self-report data through scheduled in-app surveys for daily morning EMA and weekly EPDS labels. The platform captures the following PMS data streams:
\begin{itemize}
    \item \textbf{Location:} GPS coordinates, speed, and altitude (3-minute interval).
    \item \textbf{Physical Activity:} Accelerometer data (5 Hz frequency).
    \item \textbf{Device Interaction:} Screen on/off and unlock events (real-time logging).
    \item \textbf{Social Interaction:} Incoming/outgoing call events and duration (real-time logging).
    \item \textbf{Movement States:} Activity recognition (e.g., walking, stationary) (1-minute interval).
    \item \textbf{Fitness and Status:} Step counts and battery status (5-minute interval).
\end{itemize}

Daily self-report data were collected around 8:00 AM regarding sleep quality and mood. Mood was quantified using the Circumplex Model of Affect (CMA)~\cite{Russell1980} on a 9-point Likert scale. The Three-Item-EMA used in our analyses required approximately 30 seconds, keeping the daily demand light.
Weekly EPDS surveys~\cite{epds} provided the ground truth for PPD risk. Appendix~\ref{app:data_collection_app} provides details of the user interface and system architectures of the data collection application.

\subsection{Participants and Procedure}
A total of 61 postpartum mothers in Japan (aged 20--44) were enrolled. Inclusion criteria required participants to be aged 20 or older, to own a compatible iOS smartphone, to have a first-born infant, and to provide informed consent. Baseline Social-Demographic data revealed that the infants' ages ranged from newborns to 12 months (mean: 6.8, SD: 3.3). Childcare support varied, with 90.2\% of participants supported by partners, 18.0\% living with parents, and 29.5\% utilizing nursery services. Additional details on participant demographics and recruitment are provided in Appendix~\ref{app:participant_demographics}.

Data collection occurred in three independent rounds: February 2024 (N = $18$), July 2024 (N = $18$), and September 2025 (N = $25$). Each round comprised a one-week pre-study for app verification and baseline surveys, followed by four weeks of formal data collection and final compliance-based compensation.

\subsection{EPDS Score Distribution and Participant Heterogeneity}
\label{sec:epds_distribution}
The longitudinal collection yielded 227 valid EPDS scores with a high compliance rate, and the overall missing rate for EPDS surveys was 6.97\%. Across all participants, the mean EPDS score was 6.72 ($SD=6.39$), with 26.2\% of total responses above the PPD cutoff line. Following validated clinical guidelines for the Japanese population \cite{epds_trans, epds_japanese}, we adopted a cutoff score of $\ge 9$ to define binary PPD risk.

Variance component analysis using LMM confirms that EPDS fluctuations are heavily driven by individual differences rather than temporal trends. The intercept-only model, estimated via Restricted Maximum Likelihood, revealed a high Intraclass Correlation Coefficient (ICC = 0.832). This high ICC indicates that 83.2\% of the total variance is attributable to between-user differences. Such significant heterogeneity underscores the complexity of modeling depressive symptoms across a diverse postpartum population.

\begin{figure}[tb]
    \centering
    \includegraphics[width=0.7\linewidth]{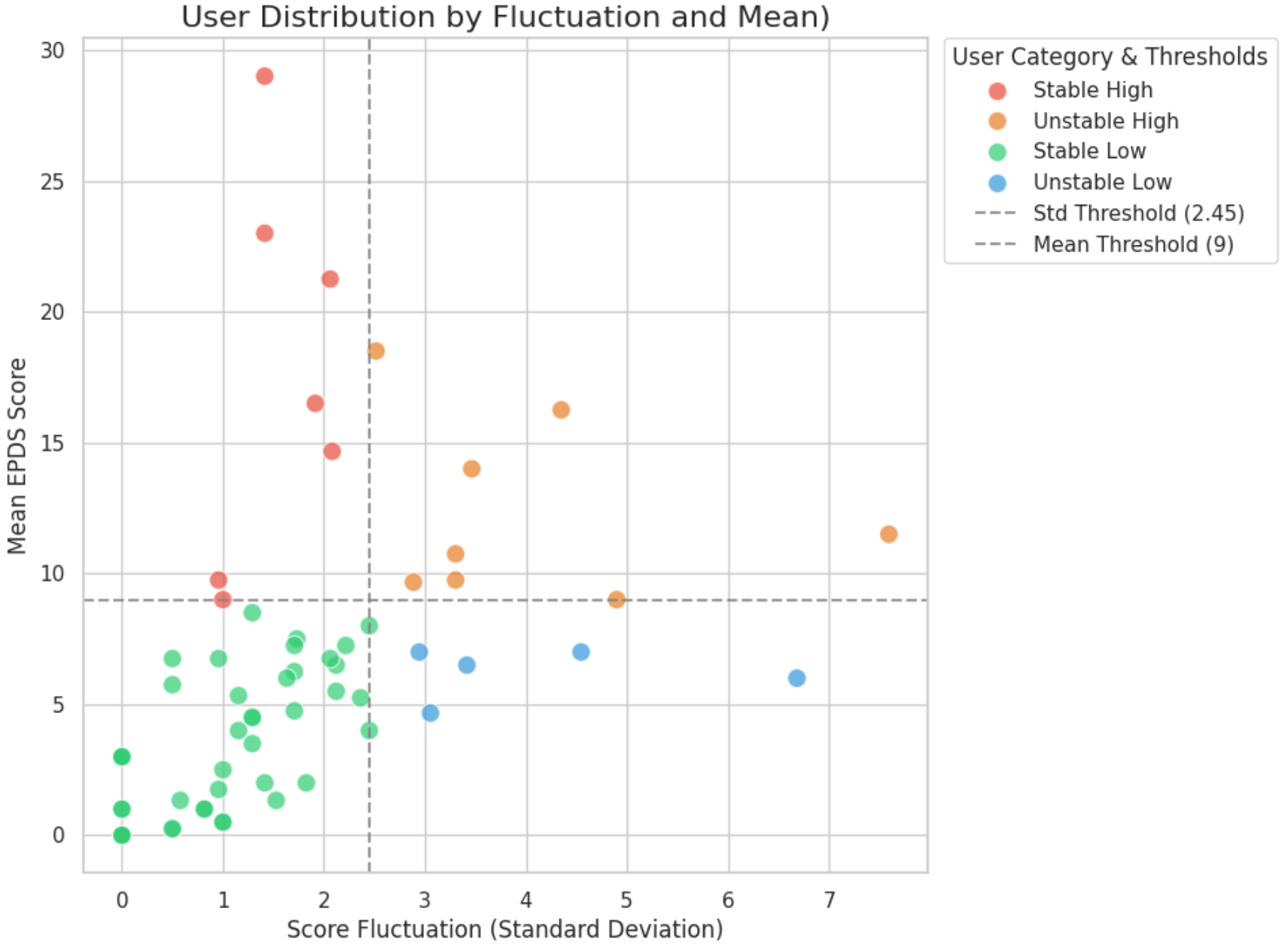}
    \caption{Participant categorization. Intra-individual fluctuations range from zero (most stable) to 7.59 (most unstable)}
    \label{fig:placeholder}
\end{figure}

To conceptualize this heterogeneity, we categorized participants into four distinct groups based on their symptom severity and stability. We utilized a mean EPDS threshold of 9 and a standard deviation threshold of 2.45 to categorize Stable Low, Stable High, Unstable Low, and Unstable High groups (Figure \ref{fig:placeholder}).
The deviation threshold isolates the top and bottom quartiles as the most unstable and stable users, respectively.
While the majority of participants (67.21\%) fall into the Stable Low category, the presence of unstable groups presents a significant challenge for generalized modeling. This characterization establishes a population-level baseline against that future personalized modeling efforts can be benchmarked.

\begin{figure}[tb]
    \centering
    \includegraphics[width=\linewidth]{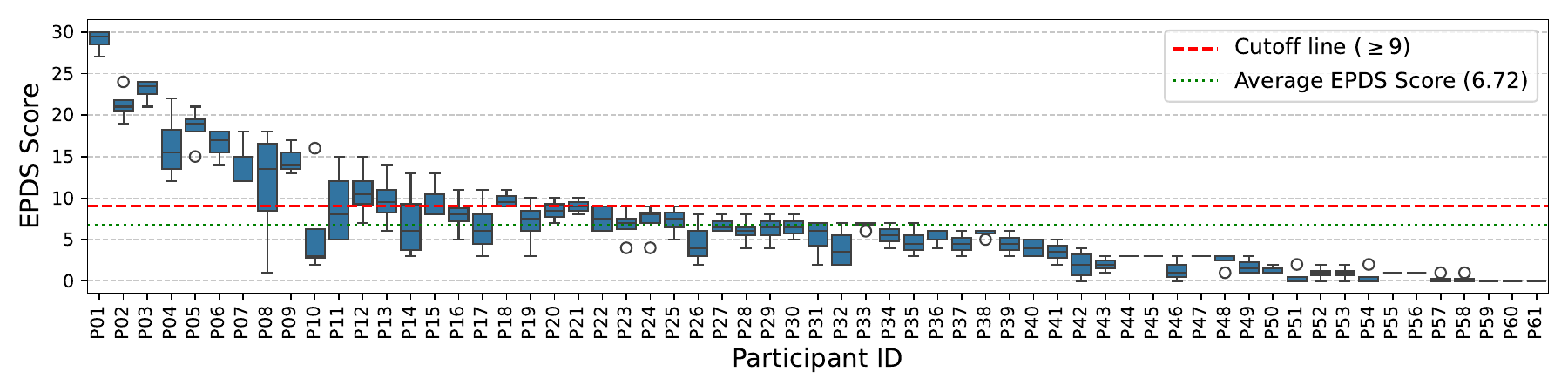}
    \caption{Distribution of EPDS scores (N=61). Mean scores for the top 10\% of users range from 16.25 to 29.00, while the bottom 10\% range from 0.00 to 0.50.}
    \label{fig:epds_distribution}
\end{figure}

\section{Method}

\begin{figure}[tb]
    \centering
    \includegraphics[width=\linewidth]{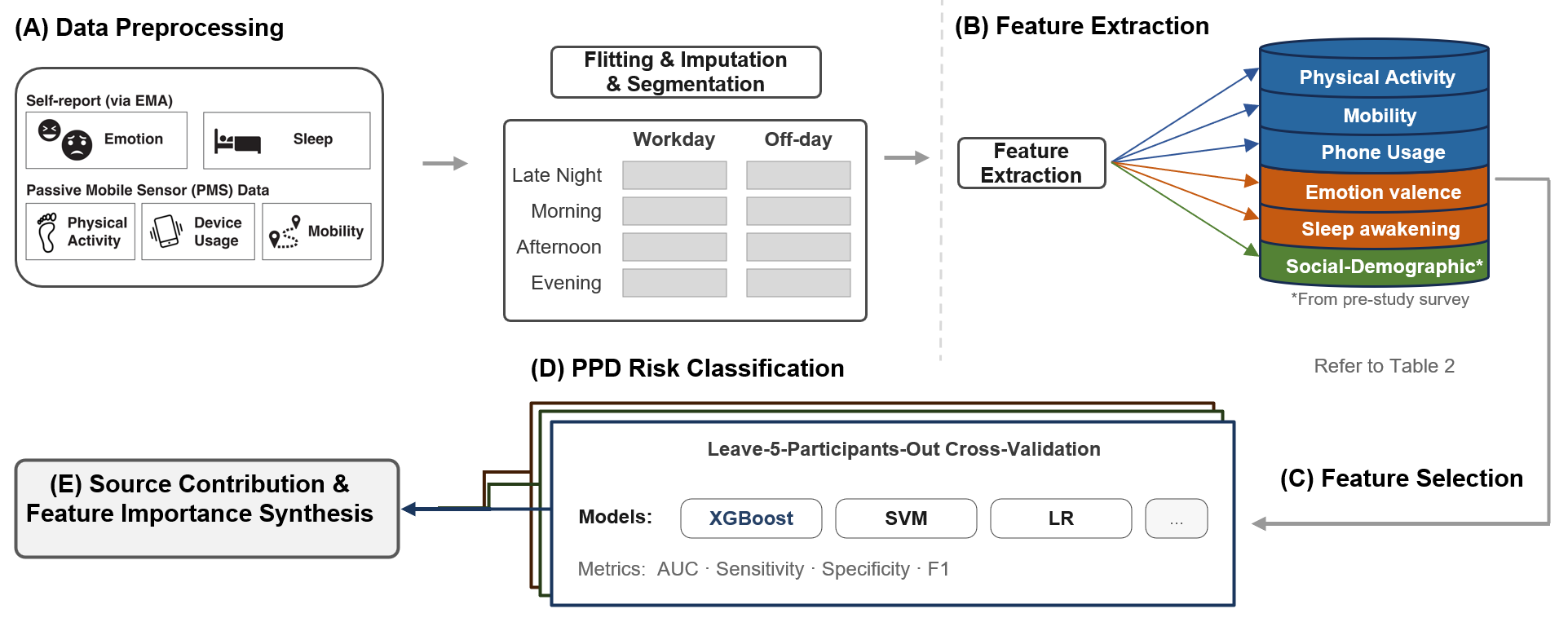}
    \caption{Data Processing and Analysis Pipeline for PPD Risk Screening.}
    \label{fig:method_flow}
\end{figure}

Our pipeline transforms raw PMS streams and self-report data—comprising daily EMA and baseline Social-Demographic surveys—into indicators of PPD risk (Figure~\ref{fig:method_flow}). The process involves multi-step data preprocessing (denoising, filtering, and segmentation), multi-modal feature extraction, and a domain-informed feature selection strategy. 
These stages feed into a participant-grouped evaluation framework to evaluate screening performance. To fulfill our research objectives, we conclude with targeted analyses to quantify source contributions (RQ1) and identify reliable digital biomarkers, including an assessment of maternal context interactions (RQ2).

Software environments, computing infrastructure, and reproducibility details are reported in Appendix~\ref{app:implementation}.

\subsection{Data Preprocessing}
\label{sec:preprocessing}

This stage prepares raw streams for feature extraction (Figure~\ref{fig:method_flow}A) through three steps: continuous-stream denoising, feature-level filtering and imputation, and a context-aware segmentation tailored to postpartum daily life.

\textbf{Continuous Stream Denoising.}
We applied accuracy-based GPS filtering to discard base-station-drift outliers and time-limited linear interpolation to bridge short gaps in continuous time-series modalities such as step counts.
These processing steps are applied exclusively to high-frequency continuous streams—specifically GPS, activity, and pedometer.
Filtering thresholds and interpolation windows are reported in Appendix~\ref{app:preprocessing}.

\textbf{Feature-Level Filtering and Imputation.}
Features with a global missing rate above 0.3 or a per-user missing rate above 0.5 were discarded. Remaining gaps were filled by user-level median imputation, with any residuals backfilled using the global median to ensure numerical stability in downstream computations. 

\textbf{Segmentation.}
We partitioned each participant's data along two axes. The temporal axis divides each 24-hour cycle into four 6-hour windows in Japan Standard Time (JST): \textit{Late Night} (00:00--06:00), \textit{Morning} (06:00--12:00), \textit{Afternoon} (12:00--18:00), and \textit{Evening} (18:00--24:00), following conventions established in prior passive-sensing depression studies~\cite{Chikersal2021}. The contextual axis labels each day as \textit{Workday} or \textit{Off-day} under a holiday-aware classification appropriate to the Japanese setting.

\subsection{Feature Extraction for Maternal Digital Biomarkers}
\label{sec:feature_extraction}

\begin{table}[tb]
  \centering
  \small
  \caption{Overview of Feature Sources and Extracted Information. All features are further aggregated across multiple temporal and contextual layers (see Appendix~ref{} for the full feature inventory).}
  \label{tab:feature_overview}
  \renewcommand{\arraystretch}{1.25}
  \begin{tabular}{l l l l p{5.8cm}}
    \toprule
    \textbf{Source} & \textbf{Feature Domain} & \textbf{Data Source} & \textbf{Sampling} & \textbf{Extracted Information} \\
    \midrule
    \multirow{5}{*}{PMS} 
      & \multirow{2}{*}{Physical Activity}
      & Pedometer       & 1/min   & step count; \textit{circadian:} IS, IV, M10 Phase \\
      & & Activity Recognition API & 1/min & active\textsuperscript{\dag} \& automotive movement  \\
    \cmidrule{2-5}
      & Mobility
      & GPS (relative coords)\textsuperscript{\ddag} & 1/10 min & displacement distance, cluster-based spatial metrics; \textit{circadian:} IS (home), IV (entropy), M10 Phase (outdoor) \\
    \cmidrule{2-5}
      & \multirow{2}{*}{Phone Usage}
      & Screen Events   & event-based & unlock count, screen-on duration; \textit{circadian:} Digital L5 Sum \\
      & & Call Logs        & event-based & call count \& duration \\
    \midrule
    \multirow{2}{*}{EMA}
      & Sleep
      & Morning EMA & 1/day & Awakenings frequency \\
    \cmidrule{2-5}
      & Emotion
      & Morning EMA   & 1/day & valence (1--5 Likert scale) \\
    \midrule
    Demo 
      & Social-demographic
      & Pre-study survey   & Once & E.g., mom/baby age, income, education level, employment status  \\
    \bottomrule
    \multicolumn{5}{p{15cm}}{\footnotesize \textsuperscript{\dag}\,Active movement is a composite indicator derived from tri-source voting (pedometer, Activity Recognition API, and GPS displacement); see text for details.} \\
    \multicolumn{5}{p{15cm}}{\footnotesize \textsuperscript{\ddag}\,GPS coordinates were converted to privacy-preserving relative coordinates (displacement from a per-user reference point) rather than raw latitude/longitude.} \\
    \multicolumn{5}{p{15cm}}{\footnotesize \textit{Circadian} metrics: IS = Interdaily Stability; IV = Intradaily Variability; M10/L5 = most/least active 10/5-hour window.} \\
  \end{tabular}
\end{table}

From the preprocessed streams, we constructed a set of digital biomarkers spanning six behavioral domains chosen to map onto the core symptom dimensions of PPD (Figure~\ref{fig:method_flow}B).
\textit{Mobility}, \textit{Phone Usage}, and \textit{Physical Activity} features were chosen to map onto dimensions of withdrawal, fatigue, and psychomotor retardation, respectively.
We supplement these PMS-derived domains with two EMA-based self-report streams---\textit{Sleep} (nocturnal awakenings) and \textit{Emotion} (momentary valence). Finally, we incorporate \textit{Social-Demographic} features (e.g., maternal and infant age, employment status, and household income) collected at the study's outset to provide necessary contextual grounding for the behavioral signals. Table~\ref{tab:feature_overview} summarizes the data sources, sampling rates, and core indicators for each domain; the full feature inventory and contextual aggregation variants are provided in Appendix~Table~\ref{tab:feature_engineering}. Phone Usage, Sleep, Emotion and Social-demographic features enter the pipeline directly from event streams or daily self-reports and require no modality-specific processing beyond the universal contextual aggregation introduced below; Mobility and Physical Activity require additional preparation, which we describe in the following paragraphs along with a set of cross-domain rhythm features.

\textbf{Multi-Layer Contextual Aggregation.}
Each raw indicator was aggregated over every cell of the temporal $\times$ day-type grid defined in Section~\ref{sec:preprocessing}, yielding a feature vector that captures behavior at varying levels of contextual specificity. This design is motivated by evidence that passive-sensing--depression associations vary markedly across time-of-day and day-type contexts~\cite{Chikersal2021}, an effect amplified for postpartum mothers whose routines are heavily shaped by infant care and work schedules. The exact aggregation grid are detailed in Appendix~\ref{app:aggregation}.
\textbf{Spatial Clustering and Exploration.}
Following established practice~\cite{Chikersal2021}, we clustered privacy-preserving relative GPS coordinates with DBSCAN and inferred each participant's home location from the most frequented Late-Night cluster. From these clusters we derived \textit{Life Space Breadth} (the daily count of distinct clusters visited), \textit{Exploration Rate} (day-to-day Jaccard dissimilarity of visited cluster sets), and segment-wise \textit{Location Distance} (cumulative Haversine displacement). Clustering parameters and metric formulas are provided in Appendix~\ref{app:spatial}.
\textbf{Activity Intensity Quantification.}
We computed two complementary movement indicators. \textit{Active Movement} is a composite intensity score that fuses pedometer counts, Activity Recognition API states, and GPS displacement, so that the resulting estimate remains robust when any single sensor fails or becomes noisy. \textit{Automotive Ratio} captures the proportion of vehicular travel within each temporal segment. The two indicators jointly distinguish self-driven mobility from motorized transport, which carry different interpretive value in the postpartum context.

\textbf{Circadian Rhythm and Routine Stability.}
To quantify the regularity and fragmentation of daily routines, we extracted four standard chronobiology metrics---Interdaily Stability (IS), Intradaily Variability (IV), and the M10/L5 windows for peak and trough activity periods. These metrics were applied to modality-specific hourly time series (e.g., step counts for physical-activity rhythms, time-at-home and spatial entropy for socio-spatial rhythms, and screen duration for digital sleep--wake patterns); definitions and per-modality applications are provided in Appendix~\ref{tab:feature_engineering}.

\subsection{Multi-Stage Feature Selection with Domain Priors}
\label{sec:feature_selection}

To distill a robust core feature set from the high-dimensional pool, we implemented a cascaded selection pipeline that compresses 364 candidates down to 17 (Figure~\ref{fig:method_flow}C). After global z-score standardization, the pipeline removes low-variance and weakly predictive features through LassoCV regularization, yielding 104 candidates. Residual collinearity is then resolved by pairing Pearson correlation thresholds with univariate linear mixed models (LMMs) that include a participant-level random intercept: whenever two features were highly correlated, the one with the smaller LMM $p$-value against the EPDS-based label was retained. The surviving candidates are then subjected to Benjamini--Hochberg FDR correction to control for multiple comparisons. 

For the final feature set, selection considered both FDR-corrected significance and domain prior knowledge. While we maintain a constant count of 17 features for all evaluated configurations to ensure comparability, the specific feature set is re-selected for each data source combination (e.g., PMS-only vs. all-sources) from the relevant feature subset. During selection, a small number of variables that prior perinatal mental health literature\cite{Krishnamurti2023, Mustafina2025} suggests may be relevant were retained alongside the statistically selected features, even when their univariate signal in this sample fell short of the FDR threshold.

The feature selection described above was applied once on the full cohort prior to cross-validation, rather than re-fitted within each fold. We discuss its rationale and limitation in 
Section~\ref{sec:limitations}.

\subsection{Classification Pipeline under Subject-wise Evaluation}

\label{sec:eval}

This subsection translates the selected feature set into trained and evaluated predictive models (Figure~\ref{fig:method_flow}D). The design is shaped by two constraints intrinsic to PMS-based PPD screening: protocols must ensure participant-wise separation between training and held-out partitions, and the resulting model must remain interpretable enough to support biomarker analyses and clinical translation.

\textbf{Task Formulation and Model Selection.}
We frame PPD risk screening as a binary classification task over weekly EPDS labels. Following the recommended Japanese cutoff~\cite{epds_japanese}, we define the primary label as $y_i = \mathbb{I}\!\left[s_i \ge 9\right]$, where $s_i$ is the participant's weekly EPDS total; an internationally used cutoff of $\ge 13$~\cite{levis2020Accuracy} is reported as a sensitivity analysis to assess robustness to threshold choice. For the model architecture, we chose four traditional estimators: Extreme Gradient Boosting (XGBoost), Random Forest (RF), Support Vector Machines (SVM), and Logistic Regression (LR). The 227-sample cohort is too small to support data-hungry deep temporal architectures without severe overfitting risk, and the tree-based and linear families chosen here allow for native explainability tools (SHAP, permutation importance) that downstream biomarker analyses depend on.

\textbf{Evaluation Protocol.}
We adopted a Leave-Five-Participants-Out (L5PO) cross-validation protocol in which the participant serves as the grouping unit, so that no individual's records appear in both the training and held-out partitions. This subject-wise split mirrors the deployment condition in which the model encounters mothers it has never seen during training, and---unlike record-wise splits---is robust to the longitudinal correlations inherent in repeated EPDS measurements. Feature-level imputation, standardization, class-imbalance reweighting, and hyperparameter selection are fitted within each training fold and only then applied to the held-out participants.

\textbf{Evaluation Metrics.}
We report AUC and Sensitivity as the primary performance metrics, as maximizing the detection of at-risk individuals is the priority in clinical screening. Specificity and F1-score are provided as supplementary indicators to assess the balance of false alarms.

\subsection{Source Contribution and Biomarker Analyses}
\label{sec:source_biomarker}

The evaluation framework above quantifies overall predictive performance, but our research questions call for two targeted analyses that build on the main pipeline (Figure~\ref{fig:method_flow}E). To address RQ1, we disentangle the contribution of PMS, EMA, and Social-Demographic sources, clarifying how much screening signal PMS carries on its own and how much each supplementary source incrementally adds. To address RQ2, we identify which specific features constitute reliable digital biomarkers by triangulating statistical and machine-learning-based evidence.

\textbf{Source Contribution Analysis.}
To quantify how much screening signal each data source carries independently and incrementally, we re-trained and re-evaluated the top-performing estimator from Section~\ref{sec:eval} on each of the three source categories \{PMS, EMA, Social-Demographic\}---comprising one full-set baseline, three leave-one-out configurations that ablate a single source, and three single-source configurations that retain only one. Every configuration uses the same L5PO cross-validation protocol, fold-internal preprocessing, and class-weight scheme as the main evaluation, and we report AUC and Sensitivity together with the change relative to the full-set baseline ($\Delta$AUC, $\Delta$Sensitivity). This factorial design jointly answers two complementary questions: whether PMS alone is sufficient for meaningful screening, and how much incremental gain each supplementary source provides on top of PMS.

\textbf{Feature Importance Synthesis.}
For each candidate feature, we triangulate two complementary lines of evidence. The first is the univariate LMM association with the EPDS-based label introduced in Section~\ref{sec:feature_selection}, which captures statistical signal under repeated longitudinal observations. The second is multivariate predictive contribution within the top-performing models: we computed mean absolute SHAP values for XGBoost (to surface non-linear contributions) and AUC-loss permutation importance for SVM and LR (to remain non-parametric for the kernel and linear estimators). The three per-model rankings were aggregated by averaging each feature's ordinal rank across estimators, yielding a single multi-model importance ranking. A feature is treated as a candidate biomarker only when both lines of evidence concur, ensuring that the surfaced markers are statistically robust at the population level and predictively influential within the multivariate decision rule.
We selected XGBoost, SVM, and LR for this synthesis as they demonstrated consistently strong discriminative performance across both the primary and sensitivity ($\ge$13) cutoffs; detailed results for the $\ge$13 threshold are provided in Appendix~\ref{sec:model13}.

\section{Evaluation and Results}

First, we establish the overall screening performance of various machine learning models using the primary EPDS cutoff of $\ge$9. This is followed by a sensitivity analysis to assess the robustness under a stricter international cutoff of $\ge$13. To futher address RQ1, we perform an ablation study to disentangle the independent and incremental contributions of PMS and self-report data sources. Finally, we investigate the underlying digital biomarkers by synthesizing statistical and machine-learning evidence, specifically examining how maternal contexts—such as employment and infant developmental stage—moderate behavioral associations(RQ2).
The discussion of how these markers diverge from general-depression phenotypes (RQ3) is deferred to Discussion~\ref{sec:discussion_biomarkers}.


\subsection{Overall Screening Performance}
\label{sec:overall_performance}

\begin{table}[tb]
\centering
\caption{Standardized Performance Metrics Comparison for the primary benchmark.}
\label{tab:model_performance_comparison}
\setlength{\tabcolsep}{5pt}
\renewcommand{\arraystretch}{1.1}

\begin{tabular}{lcccc}
\toprule
\textbf{Model} & \textbf{AUC} & \textbf{Sens.} & \textbf{Spec.} & \textbf{F1} \\
\midrule

\textbf{XGBoost}                & 0.754 ($\pm$0.153) & 0.787 ($\pm$0.299) & 0.440 ($\pm$0.256) & 0.404 ($\pm$0.168) \\

\textbf{Random Forest}          & 0.694 ($\pm$0.149) & 0.128 ($\pm$0.234) & 0.975 ($\pm$0.031) & 0.154 ($\pm$0.272) \\

\textbf{Logistic Regression}    & 0.437 ($\pm$0.225) & 0.700 ($\pm$0.458) & 0.170 ($\pm$0.222) & 0.207 ($\pm$0.167) \\

\textbf{Support Vector Machine} & 0.622 ($\pm$0.210) & 0.090 ($\pm$0.181) & 0.849 ($\pm$0.199) & 0.086 ($\pm$0.175) \\

\bottomrule
\end{tabular}

\vspace{1ex}
\RaggedRight
\footnotesize{
\textbf{Note:} 
All models utilize 17 PMS features. Values in parentheses represent standard deviations. 
Models are ranked based on the sum of their individual ranks for AUC and Sensitivity (the lower the sum, the higher the ranking).
}
\end{table}

As detailed in the Method section, we evaluated multiple classical machine learning models using exclusively PMS features to establish a baseline that reflects the core passive sensing capability of our system. Table~\ref{tab:model_performance_comparison} presents the performance metrics of all evaluated models. 
Among them, XGBoost achieved the best and most balanced performance (AUC = 0.754, Sensitivity = 0.787) and was therefore selected as the primary model for subsequent analyses.

\begin{table}[t]
\centering
\caption{Ablation Study: PPD Screening Performance with Different Feature Source Combinations}
\label{tab:ablation}
\begin{tabular}{l ccc cccc}
\toprule
\multirow{2}{*}{Combination} 
  & \multicolumn{3}{c}{Feature Sources} 
  & \multirow{2}{*}{AUC} 
  & \multirow{2}{*}{$\Delta$ AUC} 
  & \multirow{2}{*}{Sensitivity} 
  & \multirow{2}{*}{$\Delta$ Sensitivity} \\
\cmidrule(lr){2-4}
  & PMS & EMA & Demo & & & & \\
\midrule
All Sources 
  & \checkmark & \checkmark & \checkmark 
  & 0.832 {\scriptsize($\pm$0.135)} 
  & -- 
  & 0.821 {\scriptsize($\pm$0.175)} 
  & -- \\
\midrule
PMS Only
  & \checkmark & -- & -- 
  & 0.754 
  & --0.078 
  & 0.787 
  & --0.034 \\
EMA Only 
  & -- & \checkmark & -- 
  & 0.748 
  & --0.084 
  & 0.797 
  & --0.024 \\
Demo Only 
  & -- & -- & \checkmark 
  & 0.659 
  & --0.173 
  & 0.938 
  & +0.117 \\
\midrule
No Demo 
  & \checkmark & \checkmark & -- 
  & 0.784 
  & --0.048 
  & 0.707 
  & --0.114 \\
No EMA 
  & \checkmark & -- & \checkmark 
  & 0.756 
  & --0.076 
  & 0.832 
  & +0.011 \\
No PMS
  & -- & \checkmark & \checkmark 
  & 0.714 
  & --0.118 
  & 0.721 
  & --0.100 \\
\bottomrule
\end{tabular}
\vspace{2pt}

\raggedright
\noindent{\footnotesize \textbf{Note:} \textbf{PMS} = Activity/Steps + Screen Usage + Mobility; \textbf{EMA} = Emotion + Sleep;  \textbf{Demo} = Socio-demographics.  
$\Delta$ AUC and $\Delta$ Sensitivity values represent the change relative to the All Sources model.}
\end{table}

After establishing XGBoost as the optimal model, we conducted an ablation study focused on PMS to assess both its standalone screening capability and the incremental value of supplementary data sources. 
Combining all three sources yields the strongest overall performance (AUC = 0.832, Sensitivity = 0.821), indicating that active and demographic inputs serve as effective complements to PMS.
Two additional lines of evidence further confirm the primacy of PMS. Removing PMS (No PMS) causes the steepest AUC decline ($\Delta$AUC = --0.118), exceeding the impact of removing EMA ($\Delta$AUC = --0.076) or Demo ($\Delta$AUC = --0.048). Second, among single-source configurations, PMS achieves the highest AUC (0.754), outperforming both EMA (0.748) and Demo (0.659). Taken together, these results support a design philosophy in which continuous, low-burden PMS serves as the backbone of the screening system, with lightweight active assessments and demographic context layered on to maximize predictive power.

Furthermore, to examine the robustness of our findings to cutoff choice, we replicated the primary benchmark (Table~\ref{tab:model_performance_comparison}) and all-source models using EPDS\,$\ge$\,13 as an alternative risk threshold. This stricter criterion, widely adopted in international clinical settings, reduces the number of at-risk weekly observations from 60 to 32 out of 227.

Table~\ref{tab:cutoff_comparison} summarises the resulting performance shifts. Two patterns emerge. 
First, \textit{discriminative ability is largely preserved}: the AUC of the PMS-only model declines by only 0.017 (0.754\,$\to$\,0.737), and the All-Sources model declines by 0.061 (0.832\,$\to$\,0.771), indicating that the passive sensing signal identified under the Japanese-validated threshold generalizes across severity levels.
Second, \textit{sensitivity is substantially reduced under the stricter cutoff} (PMS Only: $-$0.170; All Sources: $-$0.189), accompanied by a corresponding improvement in specificity (PMS Only: 0.440\,$\to$\,0.643). This sensitivity--specificity trade-off is consistent with the meta-analytic evidence of Levis et al.~\cite{levis2020Accuracy}, who report that raising the EPDS threshold from $\ge$9 toward $\ge$13 yields higher specificity (up to 0.95) at the cost of markedly lower sensitivity (down to 0.66)---a pattern our ML models recapitulate at the cohort level.
For a screening tool where missing at-risk individuals carry greater clinical cost than false referrals, this trade-off favors the lower threshold in a limited-sample deployment. Importantly, the directional ordering of source contributions (PMS > EMA > Social-Demographic) is preserved under both cutoffs.

\begin{table}[tb]
\centering
\small
\caption{Sensitivity Analysis: Performance Shifts when Increasing EPDS Cutoff from $\ge$9 to $\ge$13.}
\label{tab:cutoff_comparison}
\setlength{\tabcolsep}{6pt}
\renewcommand{\arraystretch}{1.1}

\begin{tabular}{lcccc}
\toprule
\textbf{Configuration} & \textbf{$\Delta$ AUC} & \textbf{AUC} & \textbf{$\Delta$ Sensitivity} & \textbf{Sensitivity} \\
\midrule
XGBoost, All Sources & \textminus0.061 & 0.771 & \textminus0.189 & 0.632 \\
XGBoost, PMS Only    & \textminus0.017 & 0.737 & \textminus0.170 & 0.617 \\
\bottomrule
\end{tabular}

\vspace{1ex} 
\parbox{0.8\linewidth}{\footnotesize
  \textbf{Note:} $\Delta$ represents the change in performance metrics when the EPDS cutoff is shifted from $\ge$9 to the stricter $\ge$13. The ``AUC'' and ``Sensitivity'' columns indicate the absolute performance scores under the $\ge$13 cutoff.}
\end{table}
\subsection{Performance Variations Across EPDS Fluctuation Patterns}

To further examine how the model performs across heterogeneous user profiles, we disaggregated the prediction results by the four EPDS fluctuation pattern groups defined in Section~\ref{sec:epds_distribution}.

\begin{figure}[tb]
    \centering
    \includegraphics[width=0.7\linewidth]{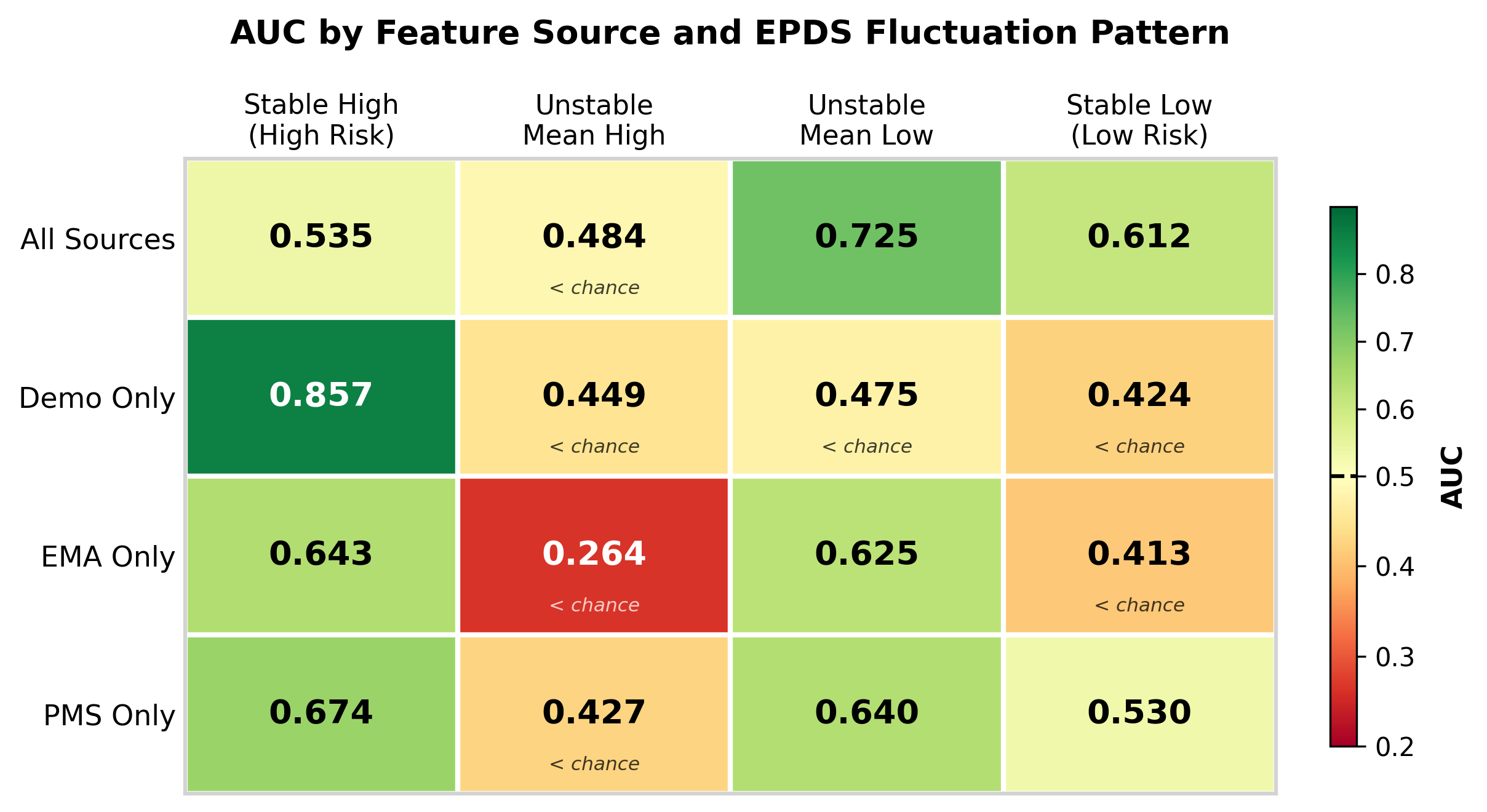}
    \caption{AUC by Feature Source and EPDS Fluctuation Pattern}
    \label{fig:auc_heatmap}
\end{figure}

It is important to note a methodological distinction: whereas the overall AUC reported in Section~\ref{sec:overall_performance} was computed as the mean of per-round AUCs in the leave-cluster-out cross-validation, the subgroup-level metrics presented here were computed by pooling predictions across all rounds to ensure sufficient sample sizes within each category.
Figure~\ref{fig:auc_heatmap} presents a heatmap of AUC values for each feature source across the four user groups, and detailed per-group metrics including sensitivity, specificity, and F1 score are provided in Appendix~\ref{tab:detailed_performance}.

Two principal findings emerge from this analysis.

\textbf{First, PMS features achieved above-chance AUC across three of the four user subgroups.}
Among the single-source models, the demographic-only model achieved the highest AUC for the Stable High group (0.857), yet its performance fell below chance level (AUC $<$ 0.50) for all three remaining groups, including Stable Low (0.424), Unstable Mean High (0.449), and Unstable Mean Low (0.475).
This pattern is theoretically expected: demographic features are collected once at enrollment and remain constant throughout the study period, rendering the model unable to track within-person symptom fluctuations.
The EMA-only model exhibited a similar brittleness, with an AUC of 0.264 for the Unstable Mean High group---the lowest observed across all model--group combinations---accompanied by a degenerate prediction pattern (sensitivity = 1.0, specificity = 0.0), indicating that the model classified all samples as positive.
In contrast, the PMS-only model maintained above-chance AUC for three of the four groups (Stable High: 0.674; Stable Low: 0.530; Unstable Mean Low: 0.640) and achieved the highest sensitivity among all single-source models for the Stable High group (0.947).
Although no single-source model excelled uniformly, PMS features exhibited the smallest variance in AUC across user types, suggesting that passively collected behavioral signals provide the most consistent predictive foundation regardless of individual symptom trajectories.

\textbf{Second, the Unstable Mean High group constituted a persistent blind spot for all models.}
As shown in Figure~\ref{fig:auc_heatmap}, AUC values for this group remained below 0.50 across all four feature configurations (range: 0.264--0.484), indicating that none of the models could distinguish high-risk from low-risk weekly observations more accurately than random assignment.
These participants---characterized by elevated mean EPDS scores with high temporal variability---represent a clinically salient subpopulation, as their fluctuating symptom severity may reflect an evolving or treatment-responsive condition that warrants close monitoring.
The consistent failure across feature sources suggests that this challenge is not attributable to the choice of input modality alone but rather reflects a fundamental limitation of population-level models in capturing highly individualized symptom dynamics.

Taken together, these findings carry two implications for future work. First, the robustness of PMS features across diverse user profiles reinforces the value of passive smartphone sensing as the core data source for postpartum depression screening, particularly in ecological settings where user compliance with active instruments (e.g., daily EMA) cannot be guaranteed.
Second, the uniformly poor performance on the Unstable Mean High group indicates that a single population-level model is insufficient to accommodate the full spectrum of individual variability observed in postpartum populations.
Future efforts should explore personalized or subgroup-adaptive modeling strategies---such as cluster-specific classifiers or online learning approaches that update predictions as individual behavioral patterns emerge over time---to address this gap.

\subsection{Digital Biomarker Identification}
To address RQ2 and investigate the specific digital biomarkers and maternal contexts that contribute most significantly to PPD risk, we executed the dual-lens analytical approach detailed in Section~\ref{sec:source_biomarker}.

Table~\ref{tab:univariate_results} presents the results from the univariate LMMs, highlighting the independent longitudinal statistical associations between individual features and PPD risk. Concurrently, the synthesized Multi-Model Rank Aggregation—derived from the machine learning interpretability techniques(SHAP and Permutation Importance)—is presented in Table~\ref{tab:feature_importance}.


\begin{table}[ht]
\centering
\caption{Univariate linearity test results}
\label{tab:univariate_results}
\begin{tabular}{lrr} 
\toprule
\textbf{Feature} & \textbf{Coef ($\beta$)} & \textbf{$P$-Value} \\
\midrule
Auto Move Ratio Evening (Mean) & 1.1419 & $0.0015^{**}$ \\
Late Night Calls (Std) & -0.5338 & $0.0053^{**}$ \\
Spouse Status & -1.9053 & $0.0081^{**}$ \\
Morning Screen Time (Std) & -0.5105 & $0.0120^{*}$ \\
Latenight Unlock Diff & 0.5179 & $0.0124^{*}$ \\
Morning Steps (Std) & 0.5949 & $0.0166^{*}$ \\
Late Night Steps (Std) & -0.7222 & $0.0197^{*}$ \\
Morning Calls (Std) & 0.5364 & $0.0204^{*}$ \\
Late Night Calls (Count) & -0.4327 & $0.0318^{*}$ \\
Valence Workday (Mean) & -0.9010 & $0.0354^{*}$ \\
Afternoon Move Score (Std) & -0.3799 & $0.0620^{\dagger}$ \\
Activity Stability & -0.5434 & $0.0868^{\dagger}$ \\
Income Rank & -1.2529 & $0.0952^{\dagger}$ \\
Morning Steps (Mean) & 0.5321 & $0.1340^{\phantom{\dagger}}$ \\
Parent Status & 1.0704 & $0.1590^{\phantom{\dagger}}$ \\
Sleep-Wake Freq (Std) & 0.2366 & $0.3430^{\phantom{\dagger}}$ \\
Baby Age & 0.6122 & $0.4160^{\phantom{\dagger}}$ \\
\bottomrule
\multicolumn{3}{l}{\footnotesize Note: 
Sorted by p-value significance.
$^{**}p<0.01$, $^{*}p<0.05$, $^{\dagger}p<0.1$. Coefs are standardized.}
\end{tabular}
\end{table}

\begin{table}[htbp]
  \centering
  \caption{Feature Importance Ranking Aggregated from Three Predictive Models}
  \label{tab:feature_importance}
  \begin{threeparttable}
    \begin{tabular}{lccr}
      \toprule
      \textbf{Feature Name} & \textbf{Top 5 Freq} & \textbf{Top 10 Freq} & \textbf{Avg Rank}  \\
      \midrule
      Sleep-Wake Freq (Std)          & 3 & 3 & 2.33  \\
      Morning Steps (Mean)           & 3 & 3 & 2.67  \\
      Late Night Steps (Std)         & 2 & 3 & 4.00  \\
      Spouse Status                  & 2 & 3 & 4.33  \\
      Valence Workday (Mean)         & 2 & 3 & 5.00  \\
      Afternoon Move Score (Std)     & 1 & 3 & 6.00  \\
      Activity Stability             & 1 & 1 & 11.00 \\
      Late Night Calls (Std)         & 1 & 1 & 11.50 \\
      Morning Steps (Std)            & 0 & 2 & 9.33  \\
      Income Rank                    & 0 & 2 & 9.67  \\
      Parent Status                  & 0 & 2 & 11.00 \\
      Latenight Unlock Diff \tnote{2}& 0 & 1 & 10.00 \\
      Auto Move Ratio Evening (Mean) \tnote{1}& 0 & 1 & 11.67 \\
      Morning Calls (Std)            & 0 & 1 & 12.33 \\
      Morning Screen Time (Std)      & 0 & 1 & 14.00 \\
      Late Night Calls (Count)       & 0 & 0 & 13.50 \\
      Baby Age                       & 0 & 0 & 14.67 \\
      \bottomrule
    \end{tabular}
    \begin{tablenotes}
      \footnotesize
      \item \textbf{Note:} Feature importance statistics are aggregated from 3 qualified models (AUC $>$ 0.7, Sensitivity $>$ 0.6) using the same 17 input features. The models include XGBoost, Logistic Regression, and SVM.
      \item \textbf{Feature Meaning:} 
      1. Represents the mean proportion of time during the evening spent in passive vehicular transport (e.g., driving or riding) as opposed to active movement.
      2. The difference of phone unlock count between workday and holiday.
    \end{tablenotes}
  \end{threeparttable}
\end{table}

Comparing LMM and ML rankings (with statistical significance defined as $p < 0.05$ in the univariate LMM) reveals three categories of features with distinct evidential profiles. \textbf{First}, a set of features attained prominence in both analyses: spousal cohabitation, late-night step variability on workdays, morning step variability, the workday--holiday late-night unlock-count difference, and workday emotional valence all reached LMM significance and ranked within the top ten of the multi-model aggregation. \textbf{Second}, sleep--wake awakening frequency variability and mean morning step count ranked first and second in the multi-model aggregation yet did not attain LMM significance; afternoon active-move-score variability and income rank followed a similar pattern, exhibiting only marginal univariate trends despite top-ten ML prominence. This divergence is consistent with the multivariate, potentially non-linear nature of their contributions: features that appear marginal in isolation may carry signal that becomes recoverable only when combined with other behavioral variables, which the LMM's univariate structure cannot capture. \textbf{Third}, evening automotive travel ratio, late-night incoming-call variability, and morning off-day screen-duration variability were each significant in the LMM but did not recur among the top-ranked features in the ML aggregation, suggesting that their associations with EPDS are identifiable in a controlled univariate setting but carry limited independent weight in the multivariate predictive context.

Beyond overall feature rankings, we conducted targeted moderation analyses on the top-ranked features identified above to test whether their associations with EPDS depend on key maternal contexts. Two robust moderation effects emerged.

\textbf{First, infant developmental stage moderated the effect of activity rhythm stability.} Although interdaily stability of activity (IS\_Activity) showed only a marginal main effect on EPDS in the univariate LMM ($\beta = -0.54$, $p = 0.087$), the IS\_Activity~$\times$~baby\_age interaction was statistically significant ($p = 0.014$). The direction indicates that the protective effect of activity stability strengthens as the infant grows: when a mother's circadian activity fails to stabilize during a period in which the infant's own routine becomes more regular, the associated PPD risk is amplified beyond what either factor predicts in isolation.

\textbf{Second, employment status moderated the effect of late-night physical activity and morning automotive mobility.} For late-night workday step features, the employment~$\times$~feature interaction was highly significant for both the mean ($p < 0.001$) and the standard deviation ($p < 0.001$): although the direction of association (positive correlation between elevated late-night activity and EPDS) held in both subgroups, the slope was steeper among unemployed mothers, suggesting that nighttime activity carries stronger risk implications in the absence of an external work schedule. By contrast, employment did not moderate morning step features (interaction $p = 0.581$ for mean, $p = 0.186$ for variability), indicating that the moderating role of employment is specific to the late-night window. A parallel pattern emerged for evening automotive ratio: while its main-effect association with EPDS was positive in the pooled sample, the employment~$\times$~ morning automotive ratio interaction was significant ($p = 0.0006$), with unemployed mothers exhibiting a substantially steeper negative slope. Together, these results indicate that employment status restructures, rather than merely shifts, the behavioral signatures of PPD risk---a pattern further unpacked in Section~\ref{sec:discussion_biomarkers}.
\section{Discussion}
\label{sec:discussion}

Our findings support the central design claim of PocketPPD: in postpartum populations, PMS carries the principal screening signal, while self-reports operate as an auxiliary modality. The remainder of the discussion is organized around three questions raised by this finding. The first subsection weighs the PMS-primary configuration against existing wearable-rich and active-data-driven approaches in terms of the performance--burden trade-off. The second unpacks which behavioral biomarkers and maternal contexts carry the screening signal, contrasting them with markers established for general depression. The third delineates the bounds of the present evidence and the directions that would consolidate it.

\subsection{Balancing Screening Performance and Burden}
\label{sec:burden_perf}

PocketPPD shows that PMS alone carries the principal PPD-screening signal, with lightweight auxiliary self-reports yielding meaningful incremental gain at modest added burden. The PMS-only configuration reaches AUC 0.75 with sensitivity 0.79; integrating EMA and Social-Demographic streams raises performance to AUC 0.83 with sensitivity 0.82. For early screening, where missed cases cost more than false positives, this sensitivity profile is favorable. Active-primary cohorts in prior PPD work report comparable or higher performance---balanced accuracy up to 93\% and AUCs of 0.82--0.84~\cite{Hahn2021, Zhong2022, Krishnamurti2023, Allen2024}---but rely on sustained self-report participation across weeks of postpartum life. This demand is difficult to maintain among mothers facing fatigue and time scarcity, and missed entries directly thin the screening signal. PocketPPD does not eliminate this trade-off but quantifies it: the source-contribution ablation enumerates configurations from PMS-only to full integration, allowing burden and performance to be weighed.

PocketPPD occupies a deliberate middle ground between wearable-rich and smartphone-coarse prior work in the PPD literature. Wearable-based models report higher physiological precision and objective sleep continuity metrics such as total sleep time and sleep fragmentation~\cite{Hurwitz2024, Slyepchenko2022, Pitsillos2022, Micheletti2020}, but introduce hardware cost and sustained-wearing barriers that limit population reach. Existing smartphone-only PPD studies, by contrast, have operated at coarse temporal resolution (e.g., monthly mobility aggregates) and reported weak or non-significant associations with depressive symptoms via classical regression rather than ML classifiers~\cite{Fransson2022}, leaving the rhythm- and volatility-based signals we recover largely unsurveyed. PocketPPD narrows this gap by engineering features at fine temporal granularity and triangulating feature importance across three classifiers (XGBoost, SVM, and logistic regression) with univariate LMM ranking.

Our reliance on classic machine learning rather than deep temporal models reflects an intentional trade-off under sample-size and clinical-transparency constraints. With 61 mothers and weekly EPDS labels, deep sequence models risk overfitting and offer limited interpretive purchase for clinical readers. Aggregating feature importance across three classifiers and triangulating with LMM significance allows meaningful indicators to surface, while leaving room for richer temporal modeling once larger longitudinal cohorts become available.

\subsection{Digital Biomarkers in the Maternal Context}
\label{sec:discussion_biomarkers}

Two themes recur across our feature-importance results: behavioral volatility and rhythm disruption dominate over absolute volume, and Social-Demographic context restructures rather than additively shifts the meaning of behavioral signals. The remainder of this subsection unpacks both themes by feature family, contrasting our findings with markers established for general depression.

\subsubsection{Social-Demographic Factors}

Unlike general depression, where Social-Demographic factors (e.g., older age and employment) directly correlate with lower baseline depression scores, Social-Demographic features in PPD interact intricately with unique maternal support structures.

Within our cohort, this distinction was visible across both analytic lenses. The LMM identified spousal cohabitation as a robust protective factor, whereas living with parents did not reach univariate significance yet still appeared among the top ten features in the multi-model ML aggregation. Income rank exhibited an analogous pattern: only a marginal univariate trend but a top-ten ML position. Neither maternal nor infant age showed a direct main-effect association with EPDS. The divergence between these two lenses suggests that specific social contexts associate with PPD risk through complex interactions rather than simple linear relationships, such that factors like intergenerational co-residence may introduce heterogeneous effects that become recoverable only when combined with concurrent behavioral streams. This perspective aligns with our observation that the all-source model substantially outperformed the Social-Demographic-only baseline (AUC 0.832 vs. 0.659), reinforcing that social context structures the meaning of behavioral signals rather than serving as a standalone screening indicator.

\subsubsection{Mobility}

Both general-depression and PPD studies conceptualize mobility through absolute spatial volume (e.g., maximum distance from home) and structural regularity, with elevated homestay reported as associated with subsequent rises in depression scores~\cite{mehrotra2018Using, ware2018Largescale}. Several PPD studies nonetheless caution that absolute travel distance alone does not consistently track depression severity, and that coarse smartphone GPS can be confounded by external context.

In our dataset, conventional metrics such as mobility radius and life-space entropy were not retained as indicators of PPD risk. The strongest mobility-related signal instead came from a contextually qualified feature: a higher proportion of evening automotive travel was positively associated with EPDS, contrasting with morning automotive ratio, which showed no comparable main effect. To probe whether this morning--evening asymmetry reflects an underlying contextual structure, we tested employment as a moderator on the sister feature, morning automotive ratio. The interaction was significant: among unemployed mothers, more morning automotive activity tracked lower EPDS, with a substantially steeper slope than among employed mothers, where no comparable association emerged.

We interpret evening automotive ratio as reflecting ``burden or escape`` trips---medical visits, caregiving-driven errands, or stress relief---rather than discretionary leisure mobility. The morning-side moderation by employment fits this reading: among unemployed mothers, whose daytime is otherwise less structured, morning automotive activity plausibly indexes intentional, agency-restoring outings. Among employed mothers, by contrast, morning travel is dominated by routine commuting and carries no comparable signal. Taken together, the morning--evening asymmetry is consistent with employment status restructuring---rather than merely shifting---the behavioral signature of mobility-related PPD risk: which mobility windows carry signal, and in which direction, depends on the daytime context against which they occur.

\subsubsection{Activity, Routine, and Circadian Rhythm}
\label{sec:activity_rhythm}

Prior general-depression studies establish both reduced absolute physical activity and disruption of circadian rhythms (e.g., lower interdaily stability) as core indicators~\cite{Wang2018, mehrotra2018Using}. Within PPD specifically, the highest-performing passive sensing models have largely depended on wearable-derived physiological signals such as heart rate and ambient light. Despite lacking such physiological ground truth, our smartphone-only configuration achieved an AUC of 0.754 by capturing fine-grained behavioral fluctuations.

Consistent with our central hypothesis, rhythm and contextually qualified activity features carried more screening weight than absolute activity volume. In the univariate LMM, step-count variability reached significance under multiple time-of-day slices (morning std, late-night workday std), whereas mean step count alone did not. Afternoon off-day active-move-score variability followed the same pattern: it ranked sixth in the multi-model ML aggregation despite only marginal LMM significance, reinforcing that std-type signals appear across multiple time-of-day slices rather than at any single window. Mean morning step count, in contrast, ranked second in the ML aggregation while remaining non-significant in the univariate LMM---a discrepancy we read as informative, suggesting that morning activity volume contributes to PPD screening primarily through interaction with other behavioral streams rather than as a standalone linear correlate. The opposing signs of morning and late-night step variability are also noteworthy: more variable morning step patterns tracked higher EPDS, while greater late-night step variability on workdays tracked lower EPDS, indicating that ``variability'' carries different valence depending on the diurnal context.

Targeted moderation analyses revealed two interactions that further refine these patterns. First, employment significantly moderated the late-night step features on workdays: although the direction of association held in both employed and unemployed subgroups, the slope was substantially steeper among unemployed mothers, indicating that nighttime activity carries stronger risk implications when no external work schedule structures the day. Second, infant developmental stage significantly moderated activity rhythm stability: although interdaily stability of activity (IS\_Activity) showed only a marginal main effect, its interaction with baby age was significant, and the protective effect of activity stability strengthened as the infant grew.

These moderation patterns have two domain-specific findings. First, late-night activity carries different semantic weight depending on a mother's daytime structure: in the absence of an external work schedule, nighttime movement is more likely to index sustained caregiving load (e.g., responding to infant wakings) than transient insomnia, steepening its association with depression. Second, the developmental moderation on rhythm stability aligns with maternal--infant chronobiology: as an infant's own circadian rhythm consolidates, the environmental cues available for maternal re-entrainment become more regular, so a mother whose activity rhythm fails to stabilize during this window may be increasingly out of phase with her infant's emerging routine---a misalignment that may amplify perceived caregiving burden.

Whether the magnitude and direction of these moderation effects generalize beyond a single Japanese cohort, where labor-market and childcare-support structures may differ, remains an open question for cross-cultural replication. Two activity-relevant modality gaps---absent ambient-light data and device-wearing patterns---further bound these findings; we return to modality issues in Section~\ref{sec:limitations}.

\subsubsection{Phone Usage: Screen Time and Communication}

Existing literature reports that higher depression scores correlate with shorter phone calls and fewer text messages, reflecting social withdrawal; less frequent communication-app usage and lower app-usage entropy are similarly associated with severe depression~\cite{Liu2023, tlachac2024Symptom}. PPD-specific work has additionally relied heavily on NLP-based semantic features, with frequent use of first-person singular pronouns and conflict-related language correlating with elevated risk.

Owing to privacy constraints and data availability, PocketPPD did not incorporate NLP features (e.g., keyboard logs, social-media text), nor did it analyze switching frequency or entropy across granular app categories (e.g., social, parenting, entertainment), as such categorization metadata was unavailable. What our analysis did surface was a consistent preference for stability- and time-segmentation-based phone-usage features over aggregate-volume metrics. Three temporally segmented features carried univariate LMM significance: late-night incoming-call variability, morning off-day screen-duration variability, and the workday--holiday gap in late-night unlock counts; non-temporal phone-usage indicators were not retained. The unlock-count gap is particularly telling: rather than indexing total volume, it captures whether late-night phone use differs structurally between workdays and off-days, and a larger gap tracked higher EPDS---consistent with the broader theme that routine breakdown across day-types, more than the level of any single day, carries the screening signal in this domain.

For postpartum mothers, the day is fragmented by infant-driven interruptions that vary substantially across days, weeks, and developmental stages. Aggregate volume metrics---total screen time, total call count---average over this fragmentation and obscure exactly the dimension that may carry diagnostic information. By contrast, contextual variability and structural-gap features more directly track the disrupted daily structure characteristic of the postpartum experience: how irregular morning screen routines are on off-days, how dispersed late-night call patterns are, and how unevenly late-night phone use is distributed between workdays and off-days.

\subsubsection{EMA: Mood and Sleep}


Prior PPD work using smartphone- or web-based active reporting platforms---rather than wearable-based physiological sensing---has shown that daily mood logs combined with baseline clinical scales attain balanced accuracy up to 93\% and AUCs of 0.82--0.84 for PPD screening~\cite{Hahn2021, Zhong2022, Krishnamurti2023, Allen2024}. Within this line of work, subjective sleep quality (e.g., ``feeling rested'') and prior depression history are well-established primary correlates.

PocketPPD incorporates EMA through morning mood surveys and recalls of nighttime awakenings; the EMA-only configuration reached an AUC of 0.748. Mean emotional valence on workdays was the fifth-ranked feature in the multi-model aggregation and showed a significant negative LMM association with EPDS, reproducing the established subjective-mood signal. The standard deviation of sleep-wake awakening frequency, however, ranked first in the multi-model aggregation despite a non-significant LMM main effect.

The contrast between univariate significance and multivariate importance for sleep metrics highlights the complex contribution of behavioral disruptions to PPD screening. Although subjective mood showed a significant linear association with EPDS scores, the variability of  awakening frequency ranked first in the multi-model aggregation despite lacking univariate significance. This suggests that the screening utility of sleep fragmentation emerges primarily through its interaction with concurrent behavioral and contextual features.

\subsection{Limitations and Future Work}
\label{sec:limitations}

We acknowledge several limitations that bound the interpretation of our findings and motivate concrete directions for future work.

\textbf{Sample size, representativeness, and external validity.} Our cohort of 61 postpartum mothers aligns with existing high-resolution PMS studies. However, it remains smaller than active-data-driven PPD cohorts, which typically exceed 500 participants. This restricted sample size limits statistical power, constraints subgroup analyses, and precludes deep learning approaches. Furthermore, recruitment through Yahoo!~Crowdsourcing bounds external validity by specifically selecting internet-active, iOS-using mothers in Japan. While iOS holds 60\% of the Japanese mobile market~\cite{mobile_market_share}, Android users and offline-recruited populations remain systematically under-represented. Future studies must pursue cross-cultural validation and incorporate multi-platform sensing data.

\textbf{Feature selection scope and exploratory framing.} As noted in Section~\ref{sec:feature_selection}
, our feature selection pipeline was fitted once on the full cohort rather than re-executed within each cross-validation fold. We piloted a fold-internal variant but found that, with 61 participants and 227 weekly observations, the resulting feature subsets and held-out performance varied substantially across folds, making the procedure unreliable as a basis for reporting. More robust alternatives such as nested randomized LR~\cite{Chikersal2021} presuppose substantially larger candidate pools and cohorts than our setting can support. Two methodological implications follow. First, the reported AUC and sensitivity represent in-sample-selected estimates with an upward bias. In contrast, relative comparisons across feature configurations remain internally valid due to identical selection procedures. Second, this study functions as an exploratory candidate-biomarker discovery effort. Surfaced markers, including routine volatility and employment moderation effects, serve as preliminary hypotheses. More importantly, these findings require confirmation in larger prospective cohorts capable of supporting nested selection and external validation.

\textbf{Unmeasured clinical and contextual variables.} Our protocol did not collect several variables the PPD literature identifies as core risk factors, most notably history of psychiatric conditions and pregnancy intendedness. The Social-Demographic and behavioral signals we report are therefore estimated without adjustment for two of the strongest known correlates of PPD, and their relative weight may shift once these are incorporated. We adopted EPDS $\geq 9$ as the primary cutoff (validated for the Japanese population) and reported only a brief sensitivity analysis at $\geq 13$; a more systematic examination of cutoff choice---including intermediate thresholds and clinician-confirmed diagnoses as ground truth---is left to future work.

\textbf{Observation window, sensing modality, and modeling choices.} Our four-week window captures short-term behavioral rhythms and weekly EPDS dynamics but does not span the full postpartum trajectory along which PPD symptoms emerge, fluctuate, and remit. Extending observation to 3--6 months postpartum would enable analysis of longer-term symptom trajectories and the temporal evolution of digital biomarkers as infants grow. Methodologically, our smartphone-only PMS configuration leaves two complementary sources of signal on the table. First, wearable-derived modalities---continuous physiological measurement, precise sleep staging, and device-wearing patterns---are unavailable in principle without a wrist-worn device and provide a level of physiological resolution that smartphone sensors cannot match. Second, several modalities that smartphones do support---most notably ambient light and Bluetooth-based co-presence---were not collected in this deployment due to platform-permission and protocol constraints, but would be feasible to incorporate in future iterations. Future studies should pursue richer modality coverage along both axes and revisit deep temporal modeling once larger longitudinal datasets become available.

\textbf{From population models to deployment.} The robustness of PMS features across diverse user profiles reinforces passive smartphone sensing as the central data source in ecological settings where active-instrument compliance cannot be guaranteed. At the same time, the persistent blind spot for the Unstable--Mean--High group shows that a single population-level model cannot accommodate the full spectrum of individual variability in postpartum populations. Subgroup-conditioned classifiers and online-learning approaches that adapt to emerging individual patterns are natural next steps. Beyond modeling, deployment requires user-centered design: determining the appropriate recipients, delivery channels, and timing of risk feedback, and securing richer data under strict privacy-preserving protocols suited to perinatal contexts.

\section{Conclusion}
In this paper, we present PocketPPD, a passive mobile sensing-based PPD screening application that leverages temporal and contextual feature engineering to capture disruptions in behavioral rhythms and shifts in stability associated with the maternal context. Using data collected from 61 postpartum mothers over a four-week period, our best-performing model, integrating passive smartphone sensing and lightweight self-report features, achieved an AUC of 0.83, while a PMS-only model still attained an AUC of 0.75. 
Ablation analysis suggested that passive sensing features contributed central screening signal, and feature importance analysis revealed that morning and late-night routine volatility serves as the most reliable digital biomarkers, moderated by infant developmental stage and employment status. These findings provide empirical evidence for wearable-free, low-burden PPD risk screening and lay the groundwork for continuous perinatal mental health monitoring. 
Future work should prioritize cross-cultural validation and explore personalized modeling strategies to capture the individual variability in behavioral patterns and risk profiles that population-level models cannot fully accommodate.

\bibliographystyle{ACM-Reference-Format}
\bibliography{citations}
\appendix
\section{Study Supplementary Information}

\subsection{Data Collection Application}
\label{app:data_collection_app}

The data collection application used a simple client-server architecture. The client-side application was implemented as a smartphone app based on the AWARE Framework~\cite{15aware1}. Its user interface was designed using Flutter, an open-source cross-platform UI framework. A Python (Flask)-based server received sensor and survey data from the client application, stored them in a SQL database, and ran on the Google Cloud Platform.

\begin{figure}[h]
    \centering
    \subfigure[Main Screen (Before joining a study)]{
        \includegraphics[width=0.23\linewidth]{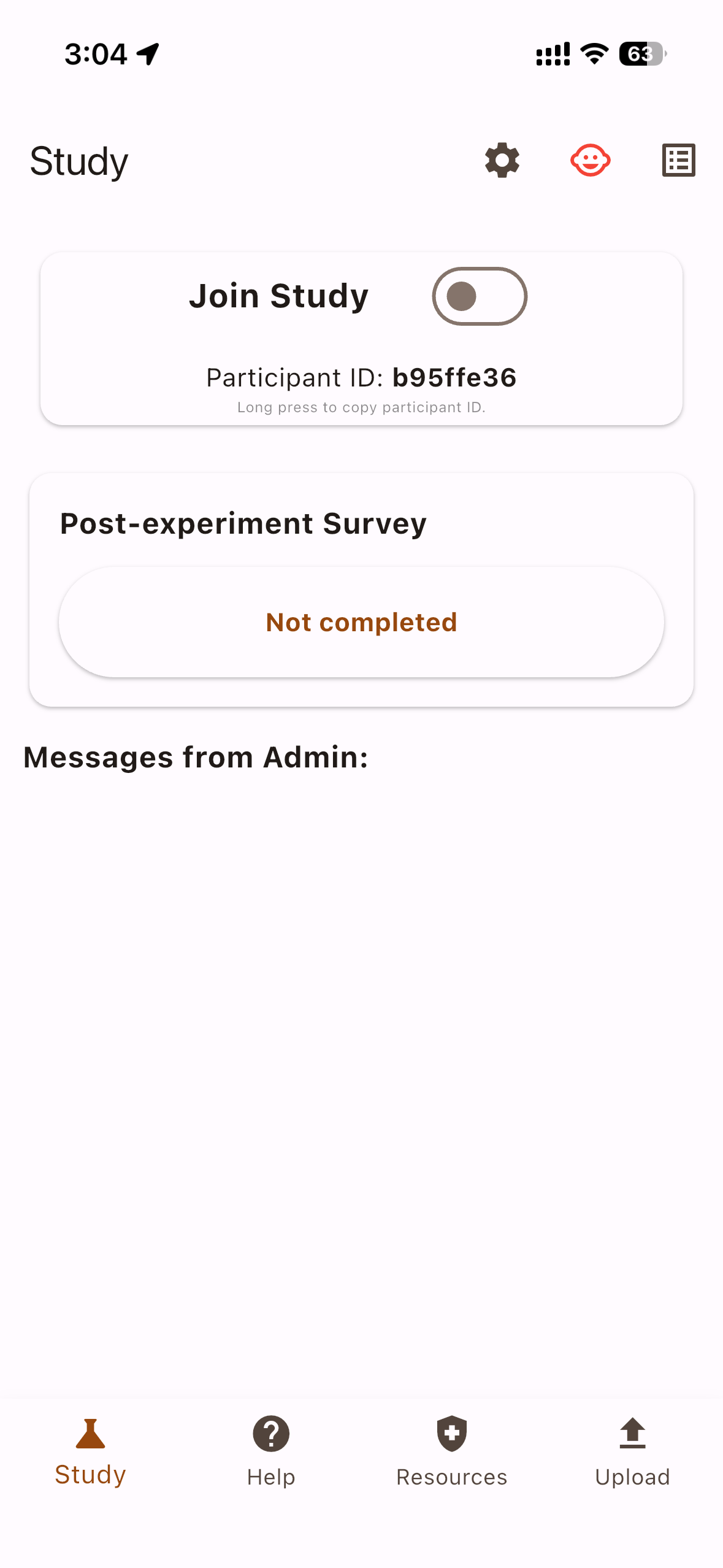}
        \label{fig:main_screen_before}
    }
    \subfigure[Settings Screen]{
        \includegraphics[width=0.23\linewidth]{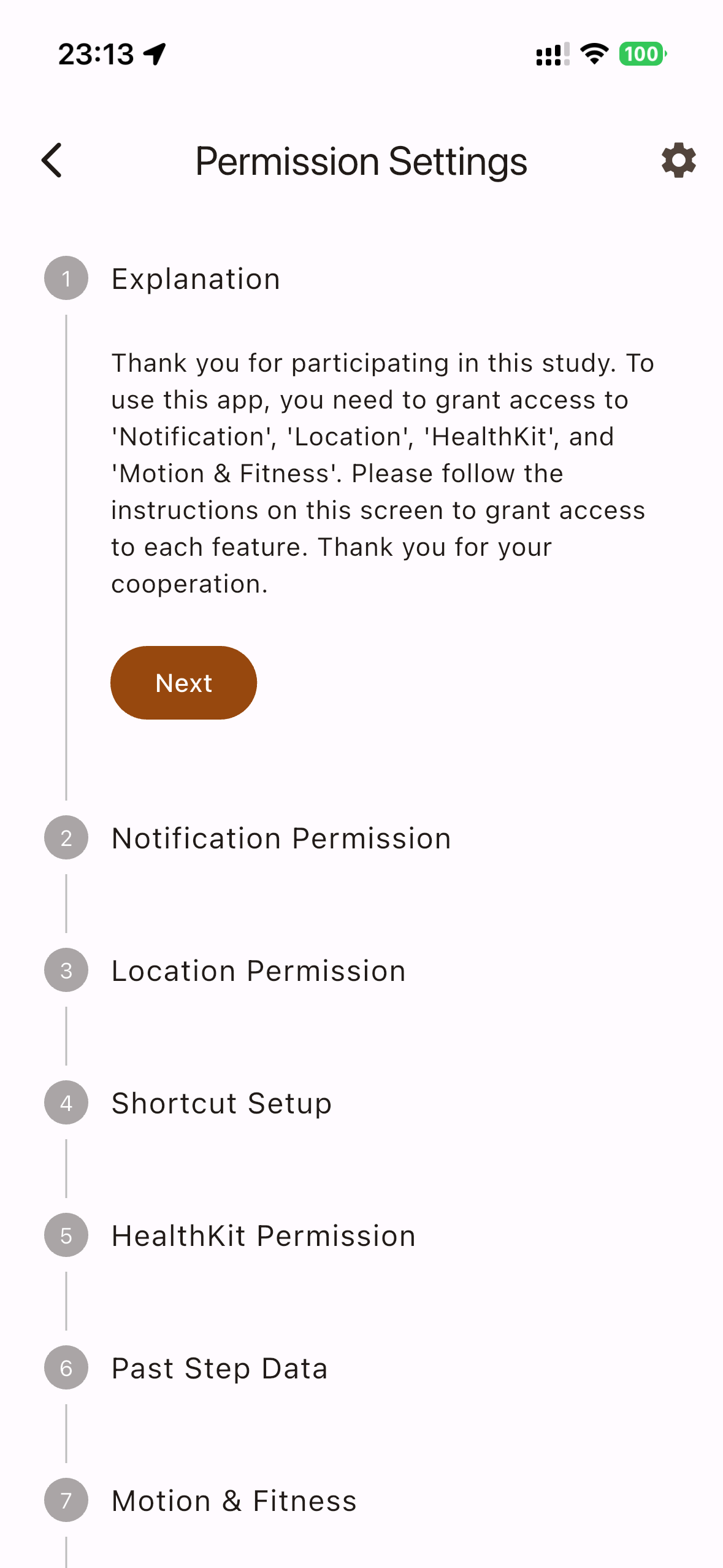}
        \label{fig:app_settings}
    }
    \subfigure[Main Screen (After joining a study)]{
        \includegraphics[width=0.23\linewidth]{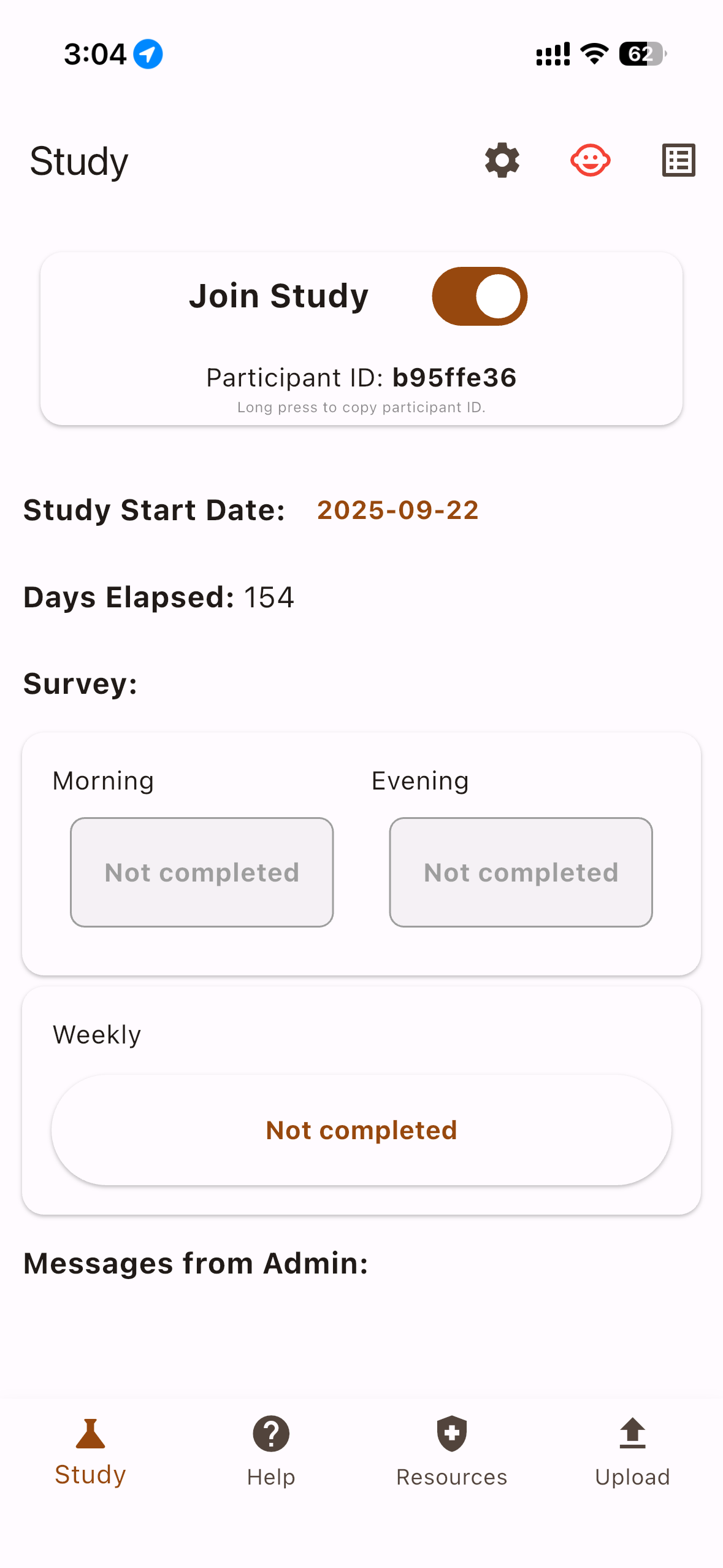}
        \label{fig:main_screen_after}
    }
    \subfigure[Survey Screen]{
        \includegraphics[width=0.23\linewidth]{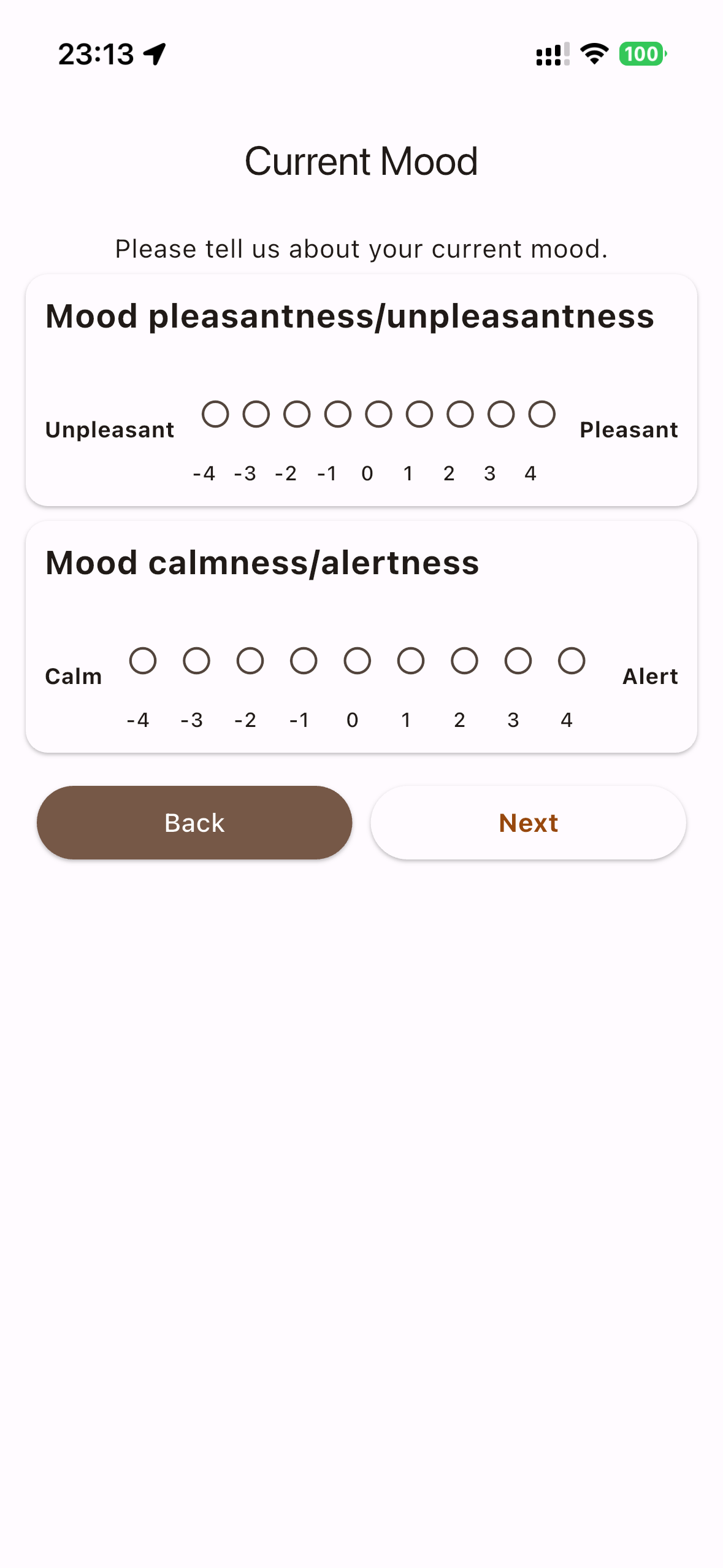}
        \label{fig:app_survey}
    }
    \caption{Screenshots of the data collection application. (a) The main screen before joining a study, where each user can see the study status and decide whether to participate via a toggle button. (b) The settings screen, where users enable access to each required sensor prior to joining. (c) The main screen after joining a study, displaying the study status including the number of days elapsed and pending in-app surveys. (d) The survey screen, where participants complete daily and weekly questionnaires including the EPDS.}
    \label{fig:app_screenshots}
\end{figure}

Figure~\ref{fig:app_screenshots} shows screenshots of the application. On the main screen, each user could see the study status and decide whether to join the study by using a toggle button. Before joining the study, each user needed to enable access to each sensor on the settings screen. After joining a study, the main screen displayed the study status, including the number of days passed and the required in-app surveys. The survey screen allowed participants to complete daily and weekly questionnaires.

\subsection{Participant Demographics}
\label{app:participant_demographics}
Participant were recruited through Yahoo Crowdsourcing \cite{YahooCrowdsourcing} for this study.

\begin{figure}[h]
    \centering
    \includegraphics[width=\linewidth]{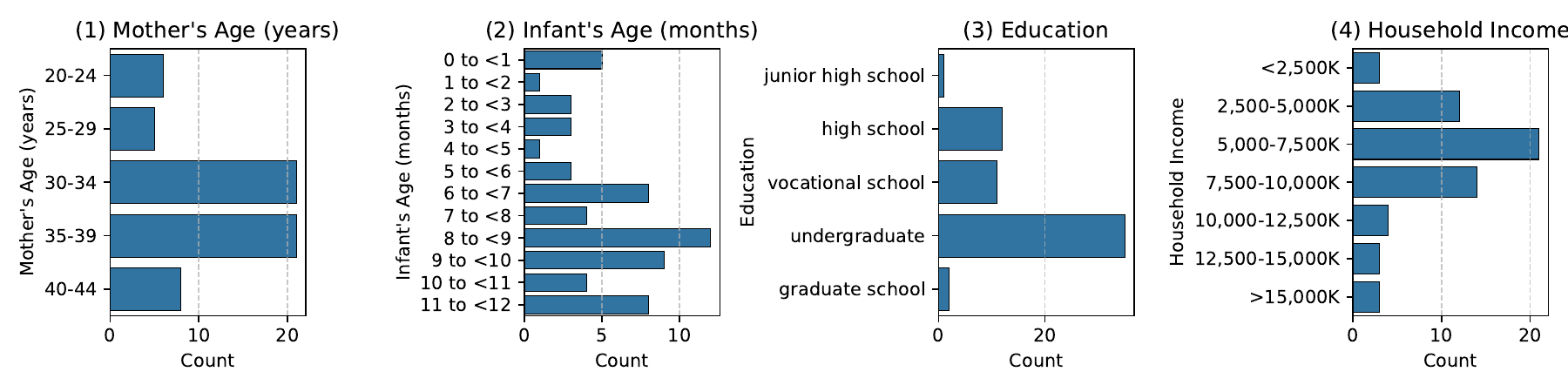}
    \caption{Age of mother and infant, education, and household income distribution of participants (N=61).}
    \label{fig:age_distribution}
\end{figure}

As shown in Figure~\ref{fig:age_distribution}, the age of mothers ranged from 20 to 44 years old, with the majority being between 30 and 39 years old. The age of infants varied from newborns to 12 months old (mean age of 6.8 months; standard deviation: 3.3). Regarding educational background, most participants had completed undergraduate education, followed by high school and vocational school/junior college. Household income levels were also diverse, with a significant portion falling within the 5,000K to 7,500K JPY range, followed by the 7,500K to 10,000K JPY range. According to the report of the Ministry of Health, Labor and Welfare in Japan~\cite{mhlw2024}, the average and median annual household income in 2023 was approximately 5,360K and 4,100K JPY, indicating that participants in our study tended to have slightly higher incomes than the population statistics; nevertheless, they appear to represent a broadly diverse range of income levels.

\section{Method Supplementary Information}

\subsection{Data Preprocessing Parameters}
\label{app:preprocessing}

As referenced in the main text, this appendix details the specific thresholds and window sizes utilized during our continuous stream denoising and data imputation processes.

\textbf{Continuous Stream Denoising Thresholds.} 
For location data, we dynamically discarded base-station-drift outliers using the GPS \textit{accuracy} metric. The filtering thresholds were set to 30m or 100m, depending on the specific environmental context and expected signal fidelity. For continuous time-series modalities such as step counts, time-limited linear interpolation was strictly limited to a window of at most two consecutive sampling points or 20 minutes. This precise constraint ensures that we smoothly bridge short gaps without artificially fabricating prolonged periods of missing data or filtering out high-frequency micro-vibration noise.

\textbf{Feature Filtering and Imputation Rules.}
To mitigate biases from missing data, we enforced strict rejection thresholds: features were discarded if their global missing rate exceeded 0.3, or if their per-user missing rate exceeded 0.5. Remaining missing segments were handled using a two-tiered strategy: user-level median imputation was applied first to preserve individual behavioral baselines, followed by global median backfilling to resolve any trace-residual missing values and ensure computational stability.

\subsection{Contextual Aggregation Grid and Circadian Metrics}
\label{app:aggregation}

As referenced in the main text, this section details the contextual aggregation grid and the mathematical formulations for the extracted circadian rhythm features.

\textbf{Contextual Aggregation Grid.}
Each day was strictly partitioned into four 6-hour diurnal windows based on Japan Standard Time (JST): \textit{Late Night} (00:00--06:00), \textit{Morning} (06:00--12:00), \textit{Afternoon} (12:00--18:00), and \textit{Evening} (18:00--24:00). Furthermore, days were categorized as \textit{Workday} or \textit{Off-day}, utilizing the \texttt{jpholiday} library to accurately account for Japanese national holidays alongside regular weekends. 
For each feature segment, we computed summary statistics (Mean and Standard Deviation). Contextual Difference (Diff) features were calculated automatically by subtracting the off-day mean from the workday mean ($\mu_{workday} - \mu_{off-day}$) for each specific diurnal window.

\textbf{Mathematical Definitions of Circadian Metrics.}
To quantify the regularity and fragmentation of daily routines, we implemented the following rhythmic features:
\begin{itemize}
    \item \textbf{Interdaily Stability (IS):} Quantifies the degree of resemblance between daily activity patterns. It is calculated as the variance of the average daily pattern divided by the total variance across all days. Values range from 0 (Gaussian noise) to 1 (perfectly repeating rhythm).
    \item \textbf{Intradaily Variability (IV):} Measures the fragmentation of activity within a single day. It is defined as the ratio of the mean squared successive differences between adjacent hours (or 10-minute bins) to the overall variance.
    \item \textbf{M10 and L5 Phase:} The M10 and L5 metrics locate the contiguous 10-hour and 5-hour windows of maximum and minimum activity, respectively. The \textit{Phase} refers to the onset time of these windows.
    \item \textbf{Digital L5 Sum:} Specifically targets sleep-time phone usage by calculating the cumulative screen-on duration within the user's identified L5 (least active 5-hour) window.
\end{itemize}

\subsection{Spatial Clustering and Exploration Parameters}
\label{app:spatial}

To extract spatial behavior patterns while strictly preserving participant privacy, all raw GPS coordinates were first transformed into relative displacements from a per-user reference point. No absolute latitude or longitude data was used in the spatial analysis.

\textbf{DBSCAN Clustering Parameters.}
We applied the DBSCAN algorithm to cluster stationary GPS points into semantically meaningful locations. A GPS sample was considered stationary if the moving speed was $\le 0.28$ m/s. The clustering parameters were set to a maximum spatial distance (\texttt{eps}) of 50 meters and a minimum cluster size (\texttt{min\_samples}) of 5 points. The participant's \textit{Home Location} was dynamically inferred as the most frequently visited cluster during the Late Night window (00:00--06:00).

\textbf{Spatial Metric Formulas.}
Based on the clustering and relative coordinates, we computed the following longitudinal metrics:
\begin{itemize}
    \item \textbf{Location Distance:} The continuous displacement calculated as the cumulative Haversine distance between successive GPS samples within each specific temporal segment.
    \item \textbf{Life Space Breadth:} Defined as the average number of distinct spatial clusters visited per day over the participant's observation window.
    \item \textbf{Exploration Rate:} Quantifies the tendency to visit novel locations versus adhering to fixed routines. It is computed as the day-to-day dissimilarity of visited cluster sets: $1 - J(C_t, C_{t+1})$, where $J(C_t, C_{t+1})$ denotes the Jaccard index between the sets of clusters visited on consecutive days ($t$ and $t+1$).
\end{itemize}

\subsection{Features}
\label{app:features}

This section provides a comprehensive breakdown of the feature engineering process and the specific variables used in predictive modeling. It includes Table~\ref{tab:feature_engineering}, which summarizes the overall feature extraction strategies, as well as Table~\ref{tab:feature_list_part1} and Table~\ref{tab:feature_list_part2}, which systematically catalog the metadata, demographic, subjective, and passive sensing features.

\begin{table*}[h]
  \centering
  \small 
  \caption{Summary of Feature Engineering. Features are categorized into behavioral sensing (Phone, Activity, Mobility, Sleep), emotional states (EMA), and demographics. The extraction involves multi-layer strategies: basic statistical aggregation (Mean/Std/Diff) across varying temporal contexts, and advanced rhythmic/longitudinal modeling (IS/IV/Phase/Exploration).}
  \label{tab:feature_engineering}
  \renewcommand{\arraystretch}{1.3} 
  {\scriptsize
  \begin{tabular}{l p{3.8cm} p{10cm}}
    \toprule
    \textbf{Category} & \textbf{Metric Source} & \textbf{Extraction Strategy \& Features} \\
    \midrule
    
    \multirow{10}{*}{\textbf{Phone Usage}} 
      & \textbf{Calls} \newline (Incoming, Outgoing, All) 
      & \textbf{Context-aware Statistics (7 Contexts):} \newline
        Extracted across Whole Week, Day Types (Workday, Off-day), and Time of Day (Morn, Aft, Eve, LateNight). \newline
        $\bullet$ \textit{Features:} Total Count, Total Duration, Duration Std (volatility), Daily Count Std. \\
    \cmidrule{2-3}
      & \textbf{Screen Interaction} \newline (Screen Duration, Unlock Count)
      & \textbf{Segmented Statistics (5 Segments):} \newline
        Extracted across Whole Day + 4 Time Segments (Morn, Aft, Eve, LateNight). \newline
        $\bullet$ \textit{Features:} Mean, Std, and \textbf{Diff} (Workday$_{\mu}$ - Off-day$_{\mu}$). \\
    \cmidrule{2-3}
      & \textbf{Sleep-Time Interaction} \newline (Derived from Screen)
      & \textbf{Rhythmic Feature:} \newline
        $\bullet$ \textbf{Digital L5 Sum}: Total screen duration during the user's least active 5 hours. \\

    \midrule
    
    \multirow{7}{*}{\textbf{Physical Activity}} 
      & \textbf{Movement Intensity} \newline (Steps, Active Score)
      & \textbf{Segmented Statistics:} \newline
        Mean, Std, and Diff calculated for 5 time segments (Whole Day + 4 Diurnal). \\
    \cmidrule{2-3}
      & \textbf{Circadian Rhythm} \newline (10-min timeseries)
      & \textbf{Rhythmic Features:} \newline
        $\bullet$ \textbf{IS (Activity)}: Interdaily Stability of physical movement. \newline
        $\bullet$ \textbf{IV (Activity)}: Intradaily Variability (fragmentation). \newline
        $\bullet$ \textbf{M10 Phase}: Onset time of the most active 10-hour window. \\

    \midrule
    
    \multirow{9}{*}{\textbf{Mobility}} 
      & \textbf{Transport \& Distance} \newline (Location Dist., Automotive Ratio)
      & \textbf{Segmented Statistics:} \newline
        Mean, Std, and Diff calculated for 5 time segments. \\
    \cmidrule{2-3}
      & \textbf{Socio-Spatial Rhythm} \newline (1-hour timeseries)
      & \textbf{Rhythmic Features:} \newline
        $\bullet$ \textbf{IS (Home)}: Regularity of time spent at home. \newline
        $\bullet$ \textbf{IV (Entropy)}: Variability of spatial entropy. \newline
        $\bullet$ \textbf{M10 Phase (Outdoor)}: Onset time of core 10-hour outdoor period. \\
    \cmidrule{2-3}
      & \textbf{Life Space \& Exploration} \newline (GPS Clusters)
      & \textbf{Longitudinal Features:} \newline
        $\bullet$ \textbf{Exploration Rate}: Day-to-day dissimilarity (1 - Jaccard Index). \newline
        $\bullet$ \textbf{Life Space Breadth}: Avg. count of distinct clusters visited daily. \\

    \midrule

    \multirow{3}{*}{\textbf{Sleep}} 
      & \textbf{Sleep Quality} \newline (Calculated from Active Logs)
      & \textbf{Context-aware Statistics:} \newline
      Extracted across Whole Week, Workday, and Off-day. \newline
      $\bullet$ \textbf{Awakenings}: Mean and Std of \texttt{total\_awakenings\_freq}. \\

    \midrule
    
    \multirow{8}{*}{\textbf{Emotion}} 
      & \textbf{EMA Responses} \newline (Valence, Arousal)
      & \textbf{Hierarchical \& Interaction-based Aggregation:} \newline
        1. \textbf{Global}: Whole Week Mean/Std. \newline
        2. \textbf{Social}: Workday vs. Off-day (Mean, Std, \textit{Social Diff}). \newline
        3. \textbf{Diurnal}: Morning vs. Evening (Mean, Std, \textit{Diurnal Diff}). \newline
        4. \textbf{Interaction}: (Workday $\times$ Morn/Eve) and (Off-day $\times$ Morn/Eve). \newline
        \textit{*Constraint: Calculated only for weeks with $\ge$ 8 EMA responses.} \\

    \midrule
    
    \multirow{3}{*}{\textbf{Social-Demographics}} 
      & \textbf{Personal \& SES} 
      & \textbf{Static Features:} \newline
        \texttt{age\_rank}, \texttt{edu\_rank}, \texttt{is\_employed}, \texttt{income\_rank}. \\
    \cmidrule{2-3}
      & \textbf{Family \& Childcare} 
      & \textbf{Static Features:} \newline
        \texttt{baby\_age}, \texttt{has\_nursery}, \texttt{has\_husband}, \texttt{has\_parent}, \texttt{childcare\_sharing\_recoded}. \\

    \bottomrule
    \multicolumn{3}{p{16.5cm}}{\footnotesize \textit{Note 1:} \textbf{Time Segments:} Morning (06:00--12:00), Afternoon (12:00--18:00), Evening (18:00--24:00), LateNight (00:00--06:00). For EMA, 'Morning'/'Evening' are defined by prompt schedules.} \\
    \multicolumn{3}{p{16.5cm}}{\footnotesize \textit{Note 2:} \textbf{Metric Definitions:} \textbf{Diff} = (Workday Mean - Off-day Mean). \textbf{IS (Interdaily Stability)}: 0-1 score measuring pattern resemblance across days. \textbf{IV (Intradaily Variability)}: Measure of rhythm fragmentation. \textbf{M10/L5}: Windows of most/least activity.}
  \end{tabular}
  }
\end{table*}

\begin{table}[h]  
    \centering
    \caption{Feature List Part 1: Metadata, Demographics, and Subjective Metrics}
    \label{tab:feature_list_part1}
    \begin{tabular}{l l p{10cm}} 
    \toprule
    \textbf{Category} & \textbf{Sub-category} & \textbf{Variable Names} \\
    \midrule
    
    Metadata & Target & \texttt{User\_ID}, \texttt{Week}, \texttt{EPDS\_Score} \\
    \midrule
    
    Demographics & Personal & \texttt{age\_rank}, \texttt{edu\_rank}, \texttt{is\_employed}, \texttt{income\_rank} \\
    \& SES & Family & \texttt{baby\_age}, \texttt{has\_nursery}, \texttt{has\_husband}, \texttt{has\_parent}, \texttt{childcare\_sharing\_recoded} \\
    \midrule
    
    Subjective & Emotion & \texttt{valence\_weekly\_mean}, \texttt{valence\_weekly\_std}, \texttt{valence\_workday\_mean}, \texttt{valence\_offday\_mean}, \texttt{valence\_diff\_off\_work} \\
    (Valence) & & \texttt{valence\_morning\_mean}, \texttt{valence\_evening\_mean}, \texttt{valence\_diff\_eve\_morn}, \texttt{valence\_morning\_workday\_mean} \\
    \midrule
    
    Subjective & Sleep Quality & \texttt{sleep-subjective\_sleep\_quality-allday\_*} (mean/std for workday, holiday, wholeweek) \\
     (Sleep) & Duration & \texttt{sleep-sleep\_duration-allday\_*} (mean/std for workday, holiday, wholeweek) \\
     & Awakenings & \texttt{sleep-total\_awakenings\_freq-allday\_*} (mean/std for workday, holiday, wholeweek) \\
     & Regularity & \texttt{sleep-bedtime\_regularity-allday\_wholeweek\_weekly-std}, \texttt{sleep-waketime\_regularity-allday\_wholeweek\_weekly-std} \\

    \bottomrule
    \end{tabular}
    
    \footnotesize{*Note: Due to length, some sleep variable suffixes (workday/holiday/wholeweek) are abbreviated.}
\end{table}


\begin{table}[h]
    \centering
    \caption{Feature List Part 2: Passive Sensing Metrics}
    \label{tab:feature_list_part2}
    \begin{tabular}{l l p{10cm}}
    \toprule
    \textbf{Category} & \textbf{Sub-category} & \textbf{Variable Names} \\
    \midrule
    
    Physical & Steps & \texttt{steps\_weekly\_total}, \texttt{steps\_daily\_std}, \texttt{steps\_workday\_daily\_mean}, \texttt{steps\_offday\_daily\_mean}, \texttt{steps\_diff\_off\_work} \\
    Activity & Intensity & \texttt{active\_score\_weekly\_mean}, \texttt{active\_score\_morning\_workday\_mean} \\
     & Movement & \texttt{distance\_weekly\_sum} \\
    \midrule
    
    Device & Screen Time & \texttt{screen\_time\_weekly\_total}, \texttt{screen\_time\_latenight\_mean} \\
    Usage & Unlocks & \texttt{unlock\_count\_daily\_mean}, \texttt{unlock\_count\_workday\_mean}, \texttt{unlock\_count\_offday\_mean} \\ 
    \midrule
    
    Mobility & Transport & \texttt{automotive\_score\_weekly\_mean}, \texttt{automotive\_ratio\_offday\_mean} \\
    \& Context & Location & \texttt{Life\_Space\_Places}, \texttt{IS\_Home}, \texttt{M10\_Phase\_Outdoor} \\
     & Social & \texttt{participation\_total} \\
    \midrule
    
    Circadian & Rhythm & \texttt{IS\_Activity}, \texttt{IV\_Activity}, \texttt{M10\_Phase\_Activity} \\
    \& Entropy & Complexity & \texttt{IV\_Entropy}, \texttt{Exploration\_Rate},\texttt{Digital\_L5\_sum}  \\

    \bottomrule
    \end{tabular}
\end{table}

\subsection{Implementation}
\label{app:implementation}
The complete computational pipeline, from data preprocessing to model evaluation, was implemented in Python 3.8. Core data manipulation and algorithmic implementations relied heavily on the Pandas and Scikit-learn ecosystems.

To handle the high-concurrency demands of nested cross-validation and iterative imputation, all machine learning experiments were conducted on the MDX platform~\cite{mdx}. Computations were executed on a dedicated virtual machine provisioned with 128 virtual cores, hosted on a server equipped with an Intel Xeon Platinum 8,368 CPU operating at 2.4 GHz. 

\section{Evaluation and Results Supplementary Information}

\subsection{Model information}

This section reports supplementary model performance metrics that complement the subgroup-level analysis in the main text. While Figure~\ref{fig:auc_heatmap} summarizes AUC values across EPDS fluctuation patterns, Table~\ref{tab:detailed_performance} provides the corresponding positive-class ratio, sample size, sensitivity, specificity, and F1 score for each feature-source configuration. These details are included to clarify whether subgroup-level differences are driven by discriminative performance, class imbalance, or sensitivity--specificity trade-offs.

\begin{table}[h]
\centering
\caption{Detailed Model Performance by User EPDS Category}
\label{tab:detailed_performance}
\resizebox{\textwidth}{!}{
\begin{tabular}{llccccccc}
\toprule
\textbf{Model Name} & \textbf{Category} & \textbf{Positive\_Ratio} & \textbf{Total Users} & \textbf{Total Samples} & \textbf{AUC} & \textbf{Sensitivity} & \textbf{Specificity} & \textbf{F1 Score} \\
\midrule
\multirow{4}{*}{All Sources} 
 & Stable High (High Risk) & 0.8214 & 5 & 28 & 0.5348 & 0.7826 & 0.2000 & 0.8000 \\
 & Stable Low (Low Risk)   & 0.0735 & 24 & 136 & 0.6115 & 0.7000 & 0.4127 & 0.1538 \\
 & Unstable Mean High      & 0.6667 & 3 & 12 & 0.4844 & 0.3750 & 0.2500 & 0.4286 \\
 & Unstable Mean Low       & 0.1667 & 3 & 24 & 0.7250 & 0.7500 & 0.5500 & 0.3750 \\
\midrule
\multirow{4}{*}{only demo} 
 & Stable High (High Risk) & 0.8750 & 4 & 16 & 0.8571 & 1.0000 & 0.0000 & 0.9333 \\
 & Stable Low (Low Risk)   & 0.1218 & 25 & 156 & 0.4239 & 0.7895 & 0.3358 & 0.2400 \\
 & Unstable Mean High      & 0.8125 & 3 & 16 & 0.4487 & 0.6154 & 0.3333 & 0.6957 \\
 & Unstable Mean Low       & 0.1667 & 3 & 12 & 0.4750 & 1.0000 & 0.0000 & 0.2857 \\
\midrule
\multirow{4}{*}{only ema} 
 & Stable High (High Risk) & 0.8214 & 5 & 28 & 0.6435 & 0.6957 & 0.2000 & 0.7442 \\
 & Stable Low (Low Risk)   & 0.0625 & 25 & 144 & 0.4128 & 0.4444 & 0.3630 & 0.0808 \\
 & Unstable Mean High      & 0.6500 & 4 & 20 & 0.2637 & 1.0000 & 0.0000 & 0.7879 \\
 & Unstable Mean Low       & 0.2500 & 2 & 8 & 0.6250 & 0.5000 & 0.6667 & 0.4000 \\
\midrule
\multirow{4}{*}{only PMS} 
 & Stable High (High Risk) & 0.7917 & 4 & 24 & 0.6737 & 0.9474 & 0.4000 & 0.9000 \\
 & Stable Low (Low Risk)   & 0.0441 & 23 & 136 & 0.5301 & 0.6667 & 0.4846 & 0.1039 \\
 & Unstable Mean High      & 0.6000 & 4 & 20 & 0.4271 & 0.7500 & 0.1250 & 0.6429 \\
 & Unstable Mean Low       & 0.2500 & 4 & 20 & 0.6400 & 0.6000 & 0.2667 & 0.3158 \\
\bottomrule
\end{tabular}%
}
\end{table}

\subsection{Model Performance for Cutoff 13}
\label{sec:model13}

This section presents an alternative evaluation using an EPDS cutoff score of 13. It includes Table~\ref{tab:model_performance_comparison}, which compares the standardized performance metrics of different models under this specific threshold.

\begin{table}[tb]
\centering
\caption{Standardized Performance Metrics Comparison(cut off = $13$).}
\label{tab:model_performance_comparison_appen}
\setlength{\tabcolsep}{5pt}
\renewcommand{\arraystretch}{1.1}

\begin{tabular}{lcccc}
\toprule
\textbf{Model} & \textbf{AUC} & \textbf{Sens.} & \textbf{Spec.} & \textbf{F1} \\
\midrule

\textbf{XGBoost}                & 0.737 & 0.617 & 0.643 & 0.384 \\

\textbf{Logistic Regression}    & 0.437 & 0.700 & 0.170 & 0.207 \\

\textbf{Support Vector Machine} & 0.622 & 0.090 & 0.849 & 0.086 \\

\bottomrule
\end{tabular}

\vspace{1ex}
\RaggedRight
\footnotesize{
\textbf{Note:} 
All models utilize 17 PMS features. 
Models are ranked based on the sum of their individual ranks for AUC and Sensitivity (the lower the sum, the higher the ranking).
}
\end{table}

\end{document}